\documentclass[aps,preprint,onecolumn,floatfix,superscriptaddress,longbibliography,nofootinbib]{revtex4-1}
\usepackage{booktabs}
\usepackage{multirow, makecell}
\usepackage{dcolumn}

\AtBeginDocument{
\heavyrulewidth=.08em
\lightrulewidth=.05em
\cmidrulewidth=.03em
\belowrulesep=.65ex
\belowbottomsep=0pt
\aboverulesep=.4ex
\abovetopsep=0pt
\cmidrulesep=\doublerulesep
\cmidrulekern=.5em
\defaultaddspace=.5em
}

\newcolumntype{P}[1]{>{\centering\arraybackslash}p{#1}}
\newcolumntype{M}[1]{>{\centering\arraybackslash}m{#1}}

\newcommand{\lsim}{\mathrel{\mathop{\kern 0pt \rlap
  {\raise.2ex\hbox{$<$}}}
  \lower.9ex\hbox{\kern-.190em $\sim$}}}
\newcommand{\gsim}{\mathrel{\mathop{\kern 0pt \rlap
  {\raise.2ex\hbox{$>$}}}
  \lower.9ex\hbox{\kern-.190em $\sim$}}}

\usepackage{amsmath}
\usepackage{graphicx,bm}
\usepackage[colorlinks,citecolor=blue]{hyperref}
\usepackage[caption=false]{subfig}
\usepackage{slashed}
\usepackage{setspace}
\usepackage[hang]{footmisc}
\usepackage{lipsum}
\usepackage{listings}

\usepackage[usenames,dvipsnames]{color}

\usepackage[utf8]{inputenc}

\definecolor{MyDarkGreen}{rgb}{0.0,0.4,0.0}

\begin{document}

\title{The Application of Convolutional Neural Networks for Tomographic Reconstruction of Hyperspectral Images}

\author{Wei-Chih Huang}
\email{huang@cp3.sdu.dk}
\affiliation{CP$^3$-Origins, University of Southern Denmark, Campusvej 55 5230 Odense M, Denmark}

\author{Mads Svanborg Peters}
\email{mape@newtec.dk}
\affiliation{Newtec Engineering A/S, 5230 Odense, Denmark}
\affiliation{Mads Clausen Institute, University of Southern Denmark, Campusvej 55 5230 Odense M, Denmark}

\author{Mads Juul Ahlebaek}
\email{maahl17@student.sdu.dk}
\affiliation{Department of Physics, Chemistry and Pharmacy}

\author{Mads Toudal Frandsen}
\email{frandsen@cp3.sdu.dk}
\affiliation{CP$^3$-Origins, University of Southern Denmark, Campusvej 55 5230 Odense M, Denmark}

\author{Ren\'e Lynge Eriksen}
\email{rle@mci.sdu.dk}
\affiliation{Mads Clausen Institute, University of Southern Denmark, Campusvej 55 5230 Odense M, Denmark}

\author{Bjarke J\o{}rgensen}
\email{bjarke@newtec.dk}
\affiliation{Newtec Engineering A/S, 5230 Odense, Denmark}

\begin{abstract}
A novel method, utilizing convolutional neural networks~(CNNs), is proposed to reconstruct hyperspectral cubes from 
computed tomography imaging spectrometer~(CTIS) images.
Current reconstruction algorithms are usually subject to long reconstruction times and mediocre precision in cases of a large number of spectral channels.
The constructed CNNs deliver higher precision and shorter reconstruction time than a sparse expectation maximization algorithm. In addition, the network can handle two different types of real-world images at the same time ---
specifically ColorChecker and carrot spectral images are considered. This work paves the way toward real-time reconstruction of hyperspectral cubes from CTIS images.

\end{abstract}

\date{\today}    

\maketitle

\section{Introduction \label{sec:introduction}}
Hyperspectral imaging~(HSI)~\cite{Goetz1147}, originating in earth observations and the launch of Landsat 1, has found several applications in control and quality assurance within the food industry. The acquired spectral information enable feature recognition, e.g. spectroscopic differentiation of materials~\cite{keshava_distance_2004}, detection of foreign objects~\cite{lee_non-destructive_2017} and optical sorting~\cite{pu_recent_2015}.
Among the existing techniques of hyperspectral imaging, pushbroom~(line scan)~\cite{boldrini_hyperspectral_2012} is the most well-known. \nocite{TitlesOn}

Despite its ability to record images of moving objects in a broad range of wavelengths,
the pushbroom system also has drawbacks: Firstly, the velocity of imaged objects has to be known to high precision and any uncertainties will result in distortion of images, and secondly, the high cost of the equipment. This hinders broader applications of hyperspectral imaging.

On the other hand, the computed tomography imaging spectrometer~(CTIS)~\cite{Okamoto:91,Th,descour_computed-tomography_1995}
 is a non-scanning snapshot HSI system, that is portable, cheaper and more compact
than the pushbroom HSI system.
 It outputs
a 2-dimensional~(2-D) image of multiplexed spatio-spectral projections surrounding a direct image of the 3-D hyperspectral cube of the field stop-limited image scene.
That is, the cube is projected into a 2-D image by superimposing the cube images in a wavelength-dependent way.
CTIS has found several applications including remote sensing microbiology~\cite{Ford:01}, ophthalmology~\cite{johnson_snapshot_2007}, space surveillance and astronomy~\cite{article} as well as food and agricultural science~\cite{8809748,Douarre:20}. One specific food science application is the detection of frost damage in carrots that is not visible in standard color RGB images.
There exist other snapshot HSI techniques, such as single-shot compressive spectral imaging with a dual-disperser architecture (CASSI) \cite{gehm_single-shot_2007}, hybrid camera multispectral-video imaging system (HMVIS) \cite{cao_high_2011}, lenslet-array \cite{bodkin_snapshot_2009} and filter-on-chip imagers \cite{von_freymann_compact_2014}.

\begin{figure}[htp]
\centering
\begin{minipage}[b]{.62\textwidth}
\subfloat[]{\includegraphics[width=1\textwidth]{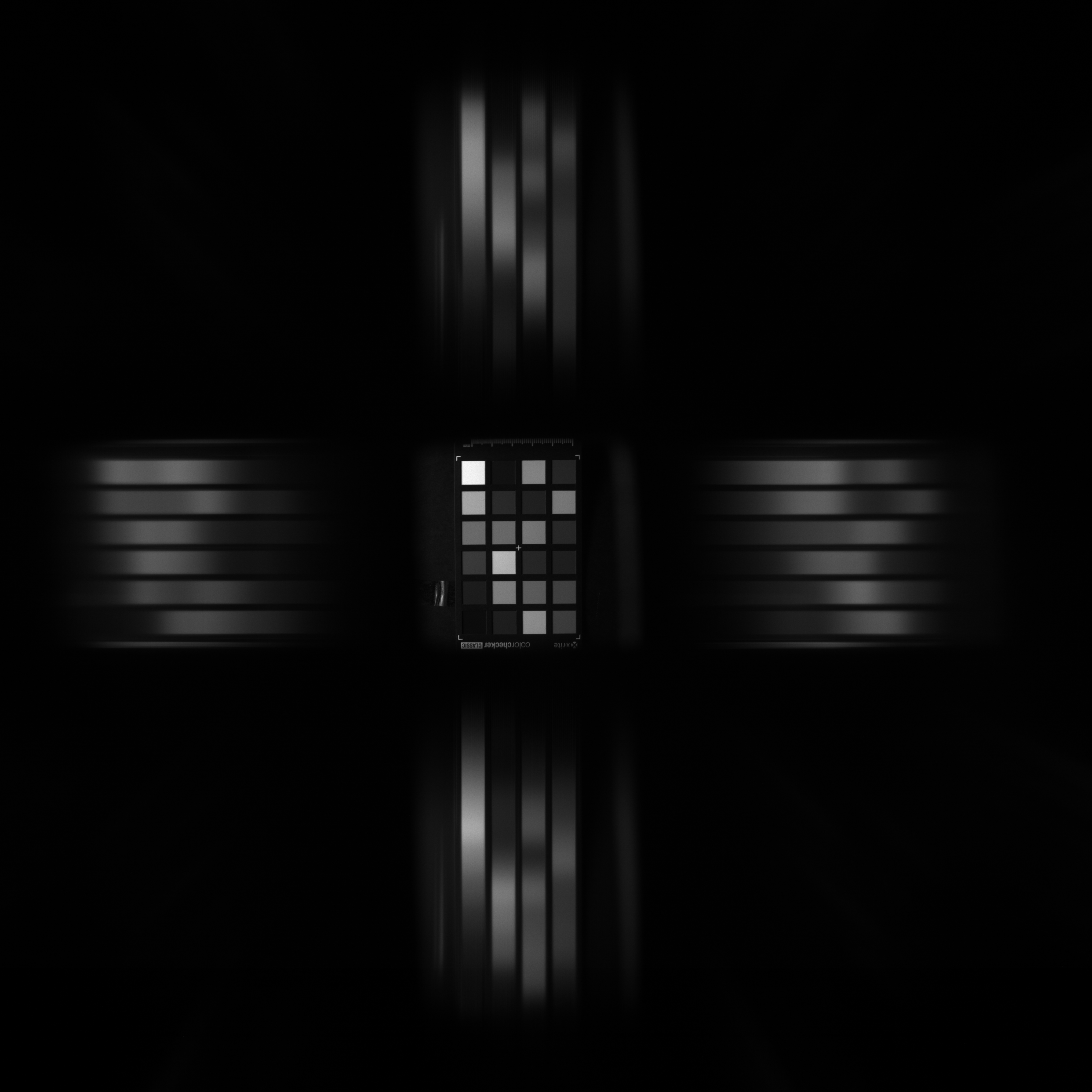}\label{Fig:rgb_0}}
\end{minipage}
\begin{minipage}[b]{.3551\textwidth}
\subfloat[]{\includegraphics[width=\textwidth]{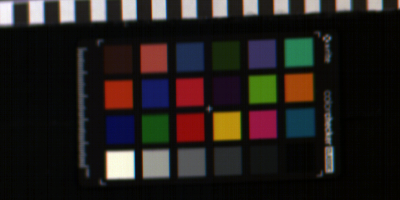}\label{Fig:rgb_a}}
\vfill
\subfloat[]{\includegraphics[width=\textwidth]{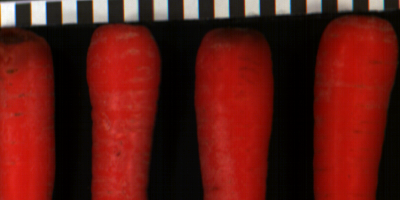}\label{Fig:rgb_b}}
\vfill
\subfloat[]{\includegraphics[width=0.67\textwidth]{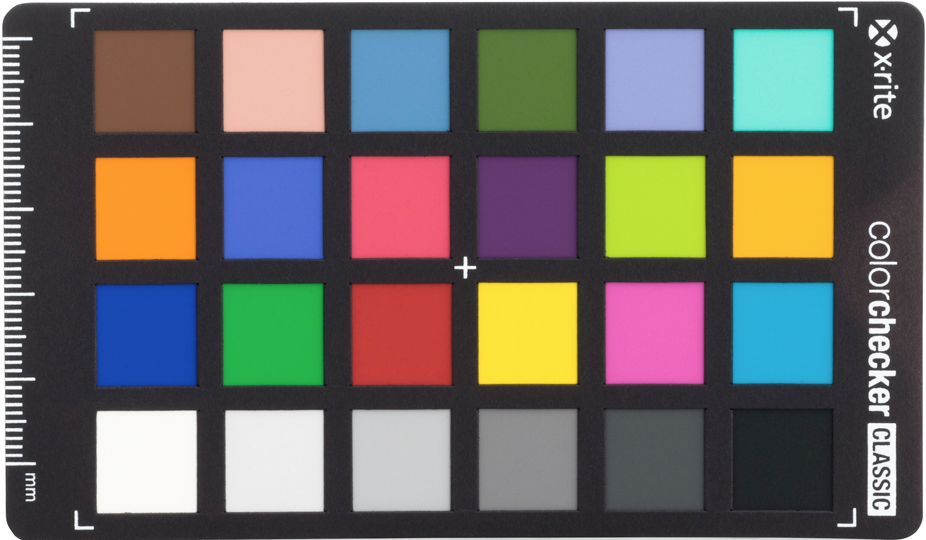}\label{Fig:rgb_c}}
\end{minipage}
\caption{(a) Experimental CTIS image of the ColorChecker acquired with a custom CTIS camera.
RGB colorization of both hyperspectral cube of (b) the ColorChecker and (c) carrots acquired  with the HSI system. (d) True RGB image of the ColorChecker. }
\label{Fig:rgb_images}
\end{figure}

In Fig.~\ref{Fig:rgb_0} we show an example of a 
CTIS image of a ColorChecker Classic Mini (ColorChecker) captured with a custom-made CTIS camera, which utilizes the same optical system layout as White et al.~\cite{white_accelerating_2020}. The CTIS image consists of a central zeroth-order
undiffracted scene image
and four surrounding first-order diffractions.
As detailed in section \ref{sec:data_CNN} 
we create synthetic CTIS images for training the neural networks 
by using a reference $200\times400\times216$ (two spatial and one spectral dimension) hyperspectral cube of the ColorChecker and carrots acquired with a pushbroom HSI system.
RGB visualizations (spectral channels 650, 550 and 470~nm) of the ColorChecker and carrots are shown in Fig.~\ref{Fig:rgb_a} and Fig.~\ref{Fig:rgb_b}, respectively, and a true RGB image of the ColorChecker is shown in Fig.~\ref{Fig:rgb_c}. 

As CTIS images are compressed and not as easy to analyze as hyperspectral cubes, 
fast and precise real-time reconstruction of the hyperspectral cube from a CTIS image is an important but challenging goal. 
In practice, the common dimension of the hyperspectral cube often exceeds $100\times100\times 100$, resulting in long reconstruction times
and mediocre accuracy for existing algorithms~\cite{vose_heuristic_2007,hagen_fourier_2007,white_accelerating_2020}. In this work, we utilize  cubes with dimensions $100\times100\times 5$ and $100\times100\times 25$: 5 spectral channels are the minimal number of channels, which achieve a substantial gain from standard RGB images, and 25 channels are a trade-off between spectral resolution and what is adequate for most applications.
Therefore, we consider neural networks to circumvent the limitations of current reconstruction algorithms. To the best of our knowledge it is the first time neural networks are used to {\it reconstruct hyperspectral cubes from CTIS images}. 

Deep neural networks~(DNNs), which can reproduce arbitrary functions given enough numbers of neurons and layers, 
are efficient at discovering underlying patterns and correlations among the input and output parameters. 
However, changing the order of input parameters leaves the output invariant --- since an exchange of two or more input features can be compensated by swapping the corresponding weights in the following hidden layer.
Therefore DNNs are not ideal for processing images where interchanging pixels of an image may alter
the image itself.
Furthermore, for high-resolution images, the resulting number of input features, proportional to the number of the image pixels, is too large to cope with for DNNs.
As a consequence, a new type of network, convolutional neural networks~(CNNs), for image processing and recognition was proposed by LeCun et al~\cite{6795724, 726791}.
Common kernels~(filters) are applied to input images for feature extraction such that one can significantly reduce the input dimensionality and capture the correlations among input features.

Recent work by Douarre et al. utilizes CNNs~\cite{Douarre:20,douarre_ctis-net_2021} to perform binary classification of apple scab lesions with various severity on apple leaves based on raw simulated CTIS images. The results are promising and outperform classification based on reconstructed cubes with standard methods.
An accompanying data set~\cite{douarre_ctis_dataset} and CTIS simulator~\cite{douarre_ctis_simulator} were also publicized. However, the dimension~($60\times 60\times 80$) of the hyperspectral cubes from the data set and CTIS simulator is different from those of our work and hence not applicable.
Additionally, our work attempts to solve the more general problem -- reconstructing the hyperspectral cube from the raw CTIS image.

In addition, there have been many applications of neural networks to image reconstruction from computed tomography~(CT) scans.
The technique combines images taken from different angles as opposed to CTIS with images from different frequencies and is used to create detailed internal images of the human body. The networks are used to, for example, reduce the image noise~\cite{Chen2017, Yang2018} or improve conventional reconstruction algorithms~\cite{2017arXiv170901809G}. In Ref.~\cite{Yang2019}, U-Net~\cite{10.1007/978-3-319-24574-4_28}, which has been extensively used for image segmentation, is employed for slice-wise reconstruction for low-dose cone-beam CT, which is related to our approach.

Since our goal is
to reconstruct a hyperspectral cube comprised of images at different wavelengths from a CTIS image, both the input and output of the network
are images. In this situation, it is natural to construct the network using only CNN layers, 
unlike tasks of image {\it classifications} where
flattening  to a 1-D vector after CNN layers and fully-connected layers are present.
To construct our  neural networks, we make use of \texttt{TensorFlow}~\cite{TensorFlow}, an end-to-end, open-source machine learning platform.
It seamlessly incorporates \texttt{Keras}~\cite{Keras}, a deep learning application programming interface written in \texttt{Python}~\cite{python}.

\section{Data preparation and CNN Architecture \label{sec:data_CNN}}

In this Section, we start by detailing how the synthetic CTIS data are created and used in the network training.
Next, we elucidate how the neural network is constructed in \texttt{Keras} and the strategies employed for training, validation and testing of the networks.

\subsection{Synthetic CTIS Data Generation}
\begin{figure}[htp]
\centering
\includegraphics[width=1\textwidth]{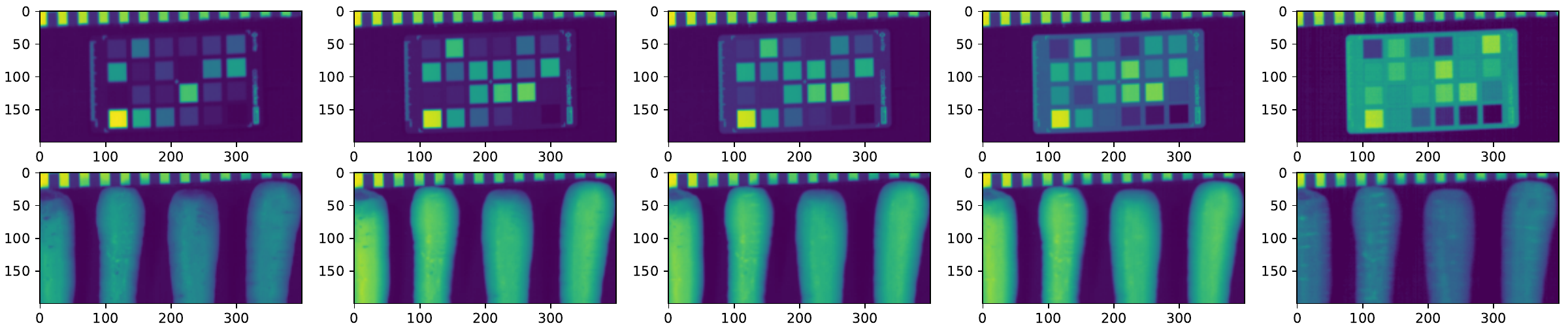}
\caption{Original spectral channels of the ColorChecker~(top row) and carrots~(bottom).
The vertical and horizontal axes indicate the location of pixels.
Each image corresponds to a single wavelength: from left to right, 603, 734, 770, 827, 969~nm, respectively.
Instead of displaying grayscale images, the colormap of  {\it viridis} is used to highlight variations in images.}
\label{Fig:5_orig}
\end{figure}
As mentioned in Sec.~\ref{sec:introduction}, we use hyperspectral cubes of the ColorChecker and carrots, acquired by the pushbroom HSI system at our disposal.
The pushbroom HSI system consists of a conveyor belt to translate the objects we image, i.e., the ColorChecker and carrots, four 150~W halogen lamps positioned above the conveyor belt and a hyperspectral pushbroom camera. The camera contains an ImSpector V10E spectrograph (Specim), a 50~mm C Series VIS-NIR objective (Edmund Optics) and a Qtechnology QT5022 computer vision system equipped with a CMOS CMV4000-E12 image sensor (CMOSIS). The pushbroom system captures 216 spectral channels ranging from 384-972~nm at a spatial resolution of approximately 0.33~mm~pixels$^{-1}$. 
For illustration, five of the 216 spectral channels are displayed in the top and bottom row of Fig.~\ref{Fig:5_orig} for the ColorChecker and carrots, respectively. We should emphasize that each image represents a specific {\it single} wavelength: from left to right, 603, 734, 770, 827, 969~nm, respectively. In other words, they are grayscale images, not standard color RGB images. 
The colormap of  {\it viridis} is, however, employed to make variations in images more pronounced.
The same colormap applies to all the following images.

\begin{figure}[htp]
\begin{center}
\subfloat[]{\includegraphics[width=0.65\textwidth]{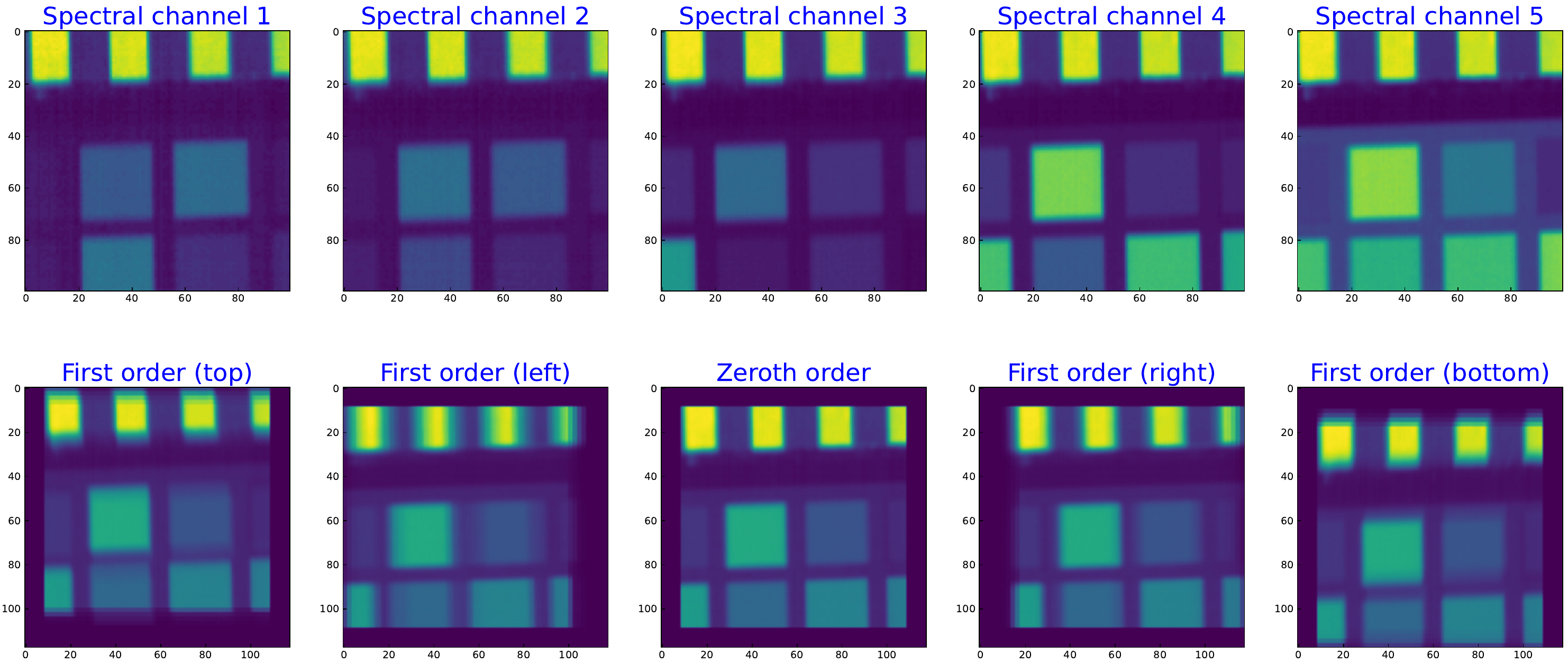}\label{Fig:5_imags_a}} \hfill
\subfloat[]{\raisebox{0ex}{\includegraphics[width=0.28\textwidth]{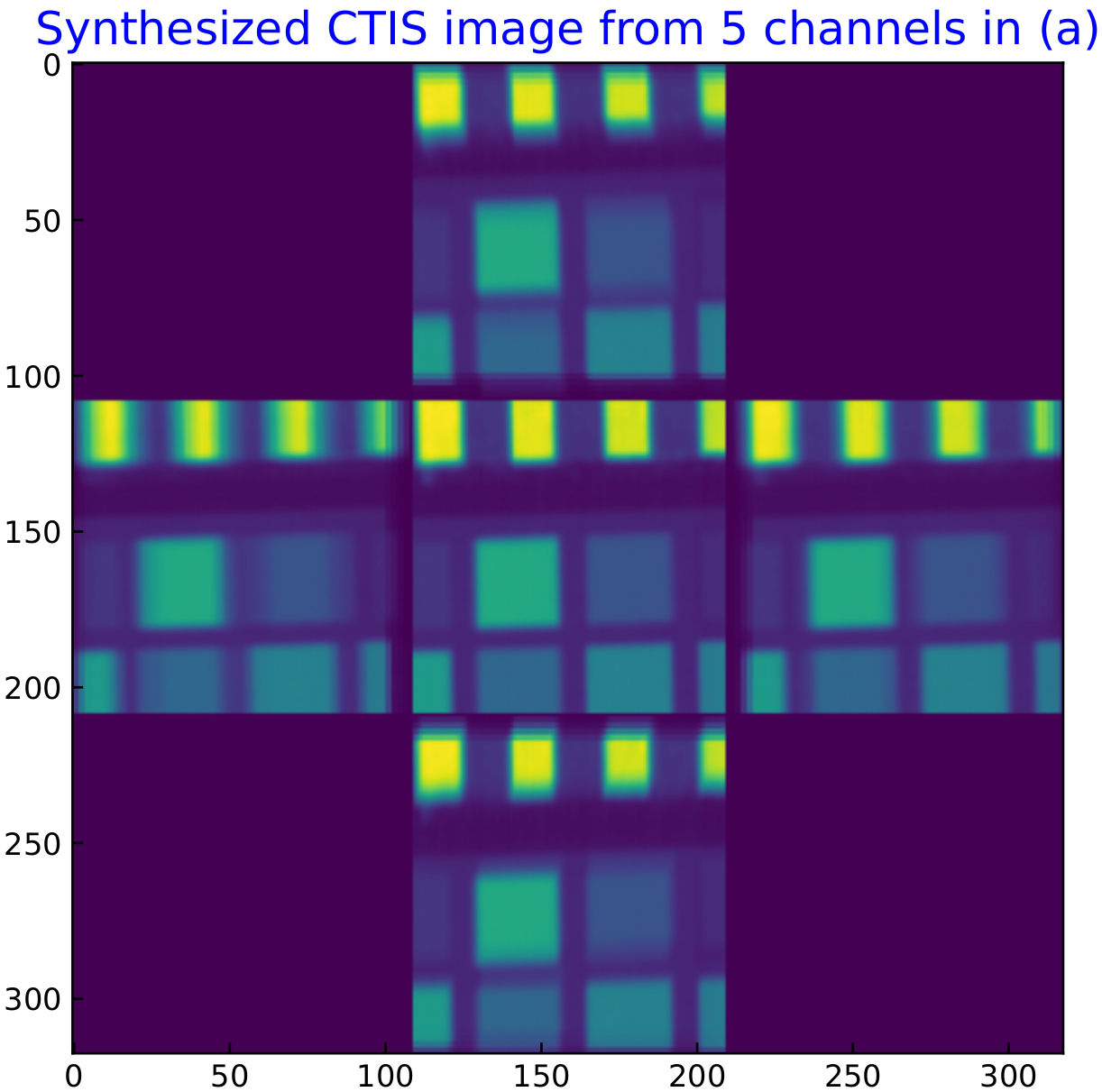}\label{Fig:5_imags_b}}}
\hfill
\caption{(a) Five spectral $100\times100$ channels of the ColorChecker (top row) with corresponding blocks of the zeroth- and first-order diffraction blocks (bottom row). (b) Synthesized CTIS image based on the five channels of the top row in (a).  }
\label{Fig:5_imags}
\end{center}
\end{figure}

To create synthetic CTIS images, we then randomly select five channels and crop each of them into smaller
$100\times100$ pixel images, resulting in 30401~(=$(200-100+1) \times (400-100+1)$) different $100\times100\times5$ hyperspectral cubes.
For each cropped hyperspectral cube, we superimpose the five channels, shown in the top row of Fig.~\ref{Fig:5_imags_a}, with the wavelengths in ascending order from the left to right.

The superimposing simulates the diffraction of the 2D diffractive optical element in the custom-made CTIS, where the channels of the hyperspectral cube are linearly dispersed  in the paraxial approximation corresponding to a wavelength-dependent pixel-shift as demonstrated in Fig.~\ref{Fig:5_imags_b}.
The central zeroth-order diffraction square in Fig.~\ref{Fig:5_imags_b}  is obtained by averaging over the five channels in the top row of Fig.~\ref{Fig:5_imags_a}.
The surrounding four first-order diffraction images are created in a similar way but with the five channels, from left to right in the top row of Fig.~\ref{Fig:5_imags_a}, shifted in a certain direction by 1, 3, 5, 7, 9 pixels, respectively, which conforms to the linear dispersion mentioned above.
The top block above the central one, for example, features upward shifts, resulting in the elongated pattern along the vertical direction.
For data preprocessing which facilitates network training, the four black corners in the resulting image (Fig.~\ref{Fig:5_imags_b}) are then removed and finally the diffraction image is divided into five block images shown in the bottom row of Fig.~\ref{Fig:5_imags_a}.
The five panels therein from left to right are simply the top, left, central, right and bottom blocks from Fig.~\ref{Fig:5_imags_b}, respectively.
Although the input hyperspectral pushbroom cubes are \textit{experimentally} acquired images (with noise), the synthetic CTIS images are simulated assuming an \textit{ideal} CTIS simulator since the current work is a proof of concept study. But future work will incorporate the optical parameters of the experimental CTIS instrument.

\begin{figure}[htp]
\begin{center}
\includegraphics[width=1\textwidth]{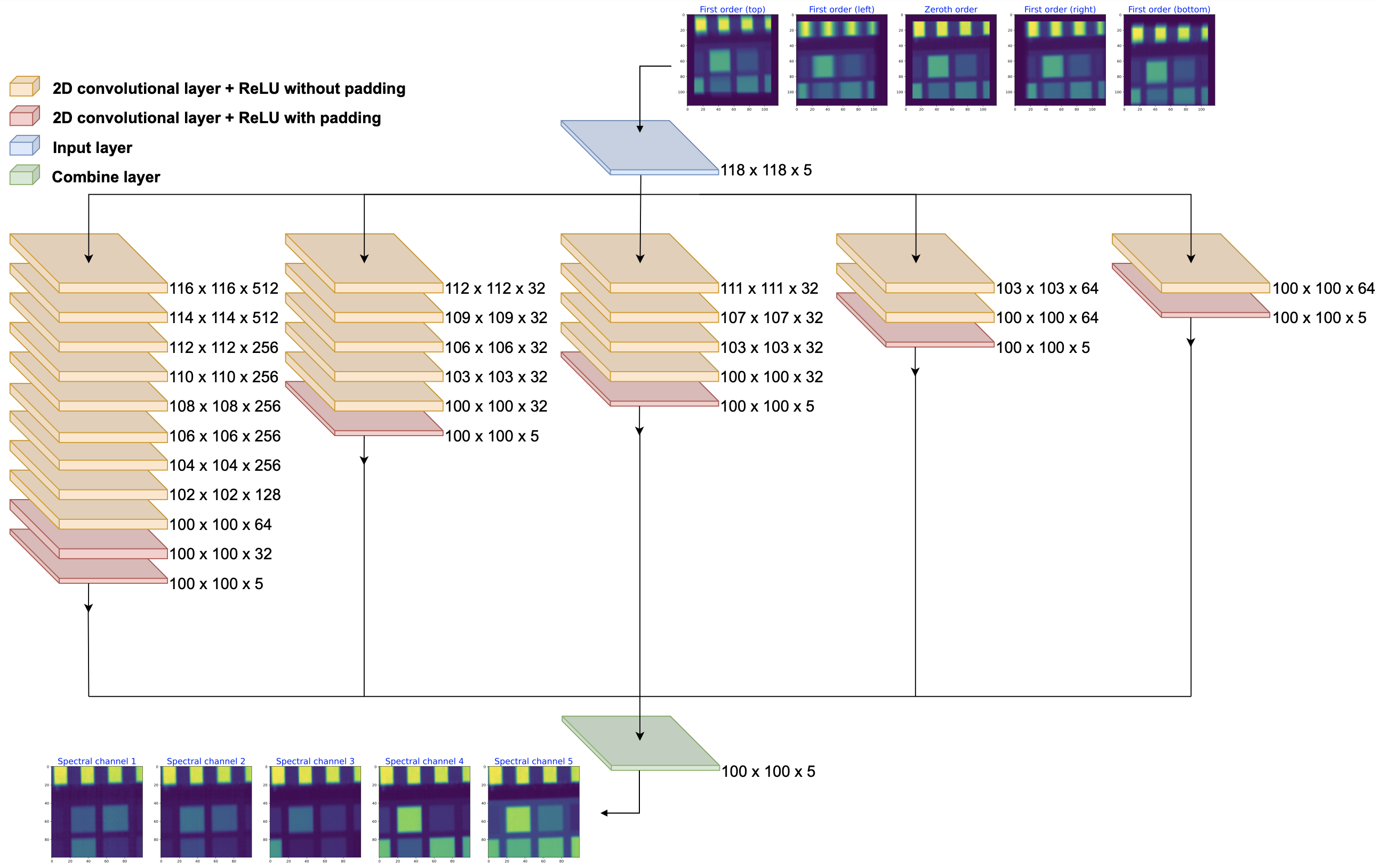}
\caption{CNN architecture for outputs of 5 spectral channels.
}
\label{Fig:CNN_5}
\end{center}
\end{figure}

Our goal is to train networks to reconstruct the original five spectral channels~(top row of Fig.~\ref{Fig:5_imags_a}) from the synthetic CTIS images~(bottom row of Fig.~\ref{Fig:5_imags_a}). 
As a result, the corresponding bottom row of Fig.~\ref{Fig:5_imags_a} comprises the input features of format (118,~118,~5), in which the first two entries in the parentheses correspond to the image dimension and the last one indicates the number of channels.
The output target variables are simply original channels~(top row) of format (100,~100,~5).

\subsection{Network architecture, data preparation and network training}
\label{sec:net_a}
Since both the input and output are images, it is natural to utilize only 2-D convolutional layers, denoted by \texttt{Conv2D} in \texttt{Keras},
without applying flattening into a 1-D array after the last layer of  \texttt{Conv2D} as usually done in CNN image classifications.
The network architecture is presented in Fig.~\ref{Fig:CNN_5}, where the top layer corresponds to the input layer and the dimensionality~(118,~118,~5)
is specified.
The input layer is followed by five distinctive branches, each composed of multiple \texttt{Conv2D} layers.
The dimensionality of their outputs is indicated; for instance, the output of the top layer in the first left branch has the dimension (116,~116,~512).
In the end, each of these five branches outputs a hyperspectral cube of dimension (100,~100,~5).
These five cubes are then concatenated and fed into the last customized output layer, dubbed \texttt{Combine layer}, which is shown at the bottom of Fig.~\ref{Fig:CNN_5}.
This layer yields a linear combination of the input cubes with bias parameters and its output dimension matches that of the cropped hyperspectral cubes.
The reason for having the five branches of different complexity is to mimic the methods of decision tree ensembles
such as random forests~\cite{ran_for} and boosted trees~\cite{10.1214/aos/1013203451} where individual trees represent
weak learners but many of them are combined to make stable predictions and achieve good performance.
In Appendix~\ref{sec:app}, we investigate how the depth of branches and the size of filters in convolutional layers influence the network performance.

For all of \texttt{Conv2D}s, 
the convolution kernel~(filter) moves one pixel rightwards or downwards on a 2-D image between two successive
applications of the kernel.
The kernel size and the number of kernels for each \texttt{Conv2D} layer can be inferred from the difference in the image dimensions between the layer's input and output. 
The first \texttt{Conv2D} in the first branch on the left, for instance, has 512 kernels of size (3,~3)~(height, width) that convert the input of (118,~118,~5) into the output of (116,~116,~512)
without padding~(without adding zeros to the boundary of the image).
For all of the last layers of the five branches, padding is used such that the dimensionality of channels is maintained.
The network contains 6.79 million trainable parameters in total.


The data involved in the training process includes 2.4 million samples\footnote{To clarify,
a sample consists of a single hyperspectral cube of $100\times100\times5$~(output) and the corresponding CTIS image~(input). }, created in three different ways: {\it full cropping}~(1.1 million), {\it sparse cropping}~(1.1 million), and {\it blank} images~(0.2 million).
\newline
The {\it full cropping} image sample is created by
taking {\it all} of 30401 possible cropped hyperspectral cubes
of size $100\times100\times5$, given a $200\times400\times5$  hyperspectral cube~(one with five specific wavelengths chosen from the original $200\times400\times216$ with 216 wavelengths).
We repeat the process for multiple randomly selected cubes of $200\times400\times5$.
Note that the process of synthesizing a CTIS image by choosing at random  5 out of 216 spectral channels is to increase the spectral variation on the training data. Therefore, the network can see more samples,
given the $200\times400\times216$ hyperspectral cube  recorded by our pushbroom system. As different spectral combinations with different frequencies have distinctive shifts instead of the fixed 1, 3, 5, 7, 9 pixels, some training samples do not correspond to physical scenarios. However, our aim is to investigate if the network can reconstruct hyperspectral cubes from a variety of CTIS images given a large amount of training data despite some of the samples being unphysical.
\newline
The {\it sparse cropping} image sample is created by
cropping only 217 cubes with a distance of 10~(15) pixels between the nearby hyperspectral cubes along the vertical~(horizontal) direction on 2-D images, i.e., strides = (10, 15) for the cropping window.
By doing so, the network is beneficially allowed to see many more different combinations
of original hyperspectral cubes without incurring an enormous dataset.
In fact, CNN has built-in partial translation invariance as a small translational shift in the input image
have little or no impact on the output of latter or deeper CNN layers\footnote{In light of dimension deduction of CNN layers without padding, a small part of the output
of a deep CNN layer corresponds to a much larger portion of the original input image.}.
As a consequence, we can afford a certain degree of sparsity in cropping.
Note that there exists no overlap of the $100\times100\times5$ cubes used in
full and sparse cropping.
\newline
Finally the {\it blank} image sample is created to avoid any bias originating from a specific image type -- the ColorChecker -- under consideration.
We include samples with only zero pixel values to mitigate any residual patterns of error associated with the properties of the training cubes.  

The 2.4 million samples are divided into the training~(2.2 million) and validation~(0.2 million) datasets.
We shall test the model on extra, unseen datasets of 1 million new cubes of $100\times100\times5$ from both full and sparse cropping.
Notice that the network has not seen the test data during training.
\newline
We use the mean squared error~(MSE) for the loss function, that quantifies the difference between the network predictions and the true images, with the mean absolute error~(MAE) and peak signal-to-noise ratio~(PSNR) in decibel  as independent metrics on errors:
\begin{align}
\text{MSE} =\frac{1}{N} \sum^N_{i=1} (Y_i - \hat{Y}_i)^2 \;\; , \;\; \text{MAE} = \frac{1}{N}  \sum_{i=1}^N  \vert Y_i - \hat{Y}_i \vert \;\; , \;\;
\text{PSNR} = 10 \log_{10} \left( \frac{255^2}{ \text{MSE} }  \right) \;\; ,
\end{align}
where $N$ is the data sample size, $i$ labels the sample,  $\hat{Y}_i$ is the network prediction and $Y_i$ is the true value. 
By comparison, MSE is more prone to outliners~(those with large residual errors) \rm{due to the quadratic dependence on the residual} while MAE tends to encourage more sparse error distributions, i.e.,
more vanishing residual errors among pixels in samples.
The Adam algorithm~\cite{2014arXiv1412.6980K}, which is a method of stochastic gradient descent based on the adaptive estimation of first-order and second-order moments, is used to minimize the loss function during training.  We choose a learning rate of $10^{-4}$.

To prevent overfitting the training data, we make use of \texttt{callbacks}, which interrupts  network training  when the loss on the validation data ceases to improve and save the layer weights with the smallest value of the loss on the validation set.
Overfitting occurs when a trained network performs very well on the data set that it was trained on but is not able to generalize to different data. It implies
that the network is overtrained and undesirably learns noise or statistical fluctuations pertaining to the training data, rather than assimilating  the underlying pattern.

\section{Results and generalizations to 25 spectral channels and carrot hyperspectral cubes
\label{sec:results_generalisation}}

In this section, we begin with presenting the training results based on the generated data discussed in Sec.~\ref{sec:data_CNN}.
Next, we take a step further to see if networks can handle a much more complicated task of decomposing CTIS images consisting of 25 $100\times100$ images, an important step
toward mimicking real-world hyperspectral cubes that usually have more than 100 spectral channels. 
Finally, we investigate if the model can manage a different type of hyperspectral cubes from synthetic CTIS images of carrots.

 \subsection{CNN predictions\label{subsec:CNN_pred} on ColorChecker images with 5 spectral channels}
We train the network on four NVIDIA Tesla V100-SXM2-32GB GPUs running in parallel. 
After 40 epochs\footnote{An epoch refers to one iteration 
where the network sees the entire training dataset once.} we reach (MSE, MAE) $\sim (0.6,~0.5)$ for the training data and $(1.0,~0.6)$ for the validation data. Given that the pixel value ranges from 0 to 255, such small residual errors indicate very good performance.
As our 
synthetic data are divided into three categories, {\it full cropping, sparse cropping and blank images}, we also calculate the average errors for each of these categories as summarized in Table~\ref{tab:error_break} .
The network performs better on the full cropping data category as there exists a significant degree of overlap among the samples  and thus less variation in this category and also due to (partial) translation invariance of CNN.

\begin{table}[htp]
	\centering
	\begin{tabular}{P{2.9cm}<{\raggedright}P{2.5cm}P{2.4cm}<{\raggedright}P{2.4cm}P{1.9cm}P{1.9cm}P{1.5cm}}
	\toprule
	Training set     & \# of samples & Test samples      & \# of samples & MSE   & MAE    & PSNR \\  \midrule
	\multirow{4}{=}{ $C_{\text{train, full }}$ + $ C_{\text{train, sparse }}$ + blank images} & $2.2\cdot10^6$ & $ C_{\text{train, full }}$  & $1\cdot10^6$ & 0.16  & 0.30 & 56.0 \\  \cmidrule{2-7} 	
	 & $2.2\cdot10^6$  & $ C_{\text{train, sparse }} $ &  $1\cdot10^6$   & 1.63   & 0.85  & 46.0 \\  \cmidrule{2-7}
	&  $2.2\cdot10^6$   & Blank~images  &  1 & $4.5 \cdot 10^{-3}$ & $5.2 \cdot 10^{-2}$ &71.6  \\   \cmidrule{2-7}
	& $2.2\cdot10^6$  & $C_{\text{test}}$ &$1\cdot10^6$   & 1.86      & 0.89 & 45.4 \\
	\bottomrule
\end{tabular}
	\caption{MSE, MAE and PSNR for different  test samples, indicated by the third column, in the scenario of 5 spectral channels. The ColorChecker is abbreviated as $C_{\text{dataset}}$: $C_{\text{train, full~(sparse) }}$ refers to the full~(sparse) cropping images from the training set while $C_{\text{test}}$ represents the test set, created by both full and sparse cropping. The number of samples for training and testing are indicated in the second and fourth column, respectively. Samples from sparse cropping have larger errors than those from full cropping, while the blank images, used to reduce the bias, have the smallest errors. 
	}
	\label{tab:error_break}
\end{table}
By contrast, samples from the sparse cropping category consist of many different combinations of the $200\times400\times5$ hyperspectral cubes, featuring much more variety so 
it becomes harder for the network to make correct predictions in this category.
On the other hand, the network attains consistent performance on the completely new CTIS images from both full and sparse cropping\footnote{Note that the test data is generated based on completely distinct hyperspectral cubes from those of the training data. That is quite different from the usual practice, where training, validation and test data originate from the same source of data.}.
The corresponding values of MSE and MAE are similar to
those of sparse cropping in the training data, demonstrating that the network generalizes well to completely new data samples.

\begin{figure}[htp]
\begin{center}
\includegraphics[width=0.85\textwidth]{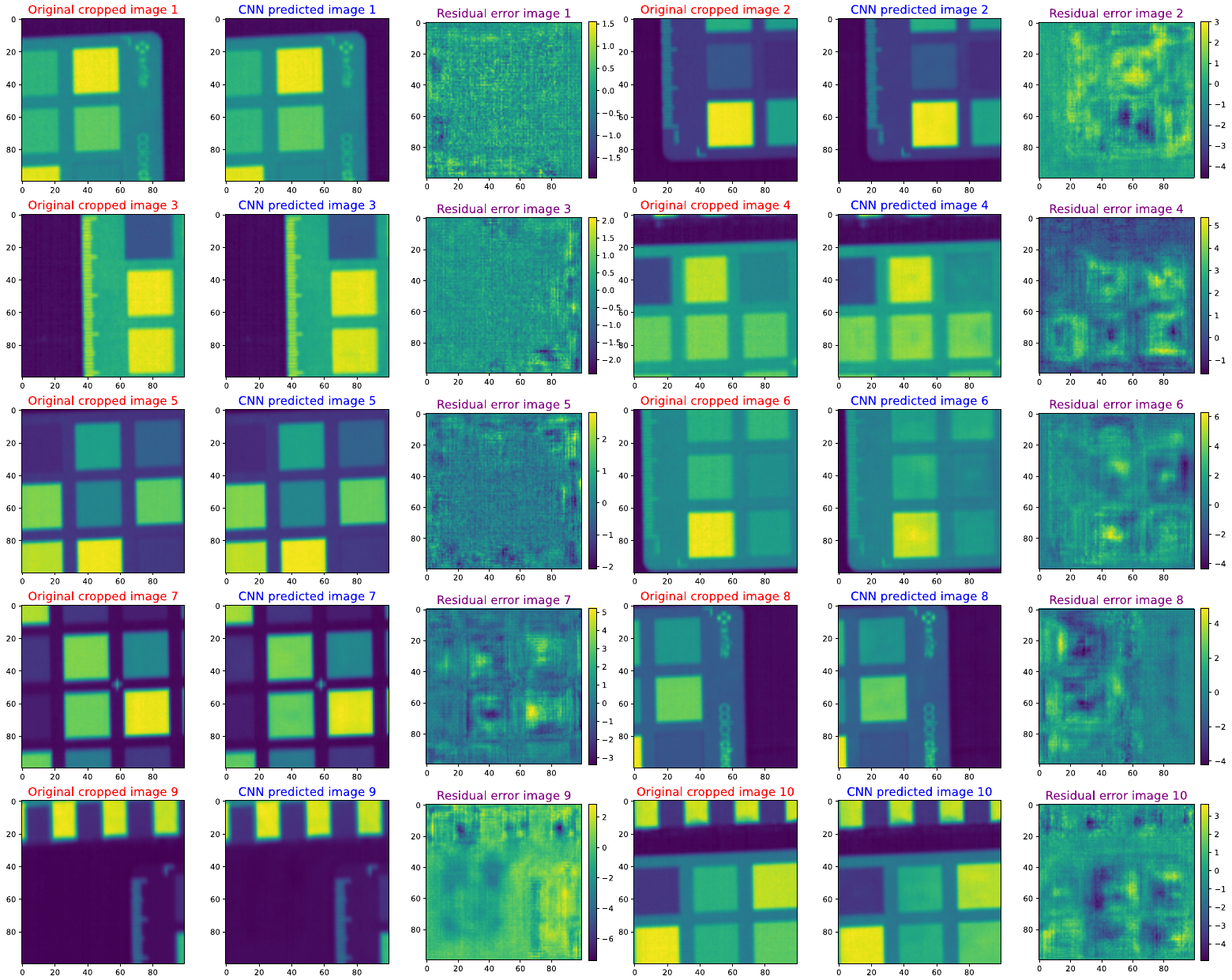}
\caption{Comparison between the original channels and CNN predictions for samples from training~(three columns on the left) and test~(three on right) datasets. Here, we show the comparison only for the third out of the five channels. For the residual error images, we add colorbars to indicate the absolute scale of errors where the ranges of the colorbars vary.
}
\label{Fig:images_5_pred}
\end{center}
\end{figure}
Fig.~\ref{Fig:images_5_pred} shows a comparison between the CNN predictions and the original images for channel 3 of the five channels used. It illustrates the quantitative results of Table~\ref{tab:error_break}.
The odd image indices (the first 3 columns from the left) label samples from the training datasets and the even indices (the last 3 columns) label samples from the test datasets. For both the training and test sets we display, from left to right, the original image of channel 3, the CNN prediction and the difference between the two.
The samples in the first three rows are from the full cropping category and the last two rows are from the sparse category. 
It is evident that the residual error in the training data is quite small for the full cropping data category~(top three panels of the third column) as compared to those of the sparse.
The larger difference between the true and predicted pixel values, the more visible the residual error pattern, which is also indicated by larger ranges of error colorbars -- for example, the residual errors for image 1 are smaller than  those of image 6.
 It is also evident that the model generalizes well to the unseen samples~(images with even indices).
\begin{figure}[htp]
\begin{center}
\subfloat[]{\includegraphics[width=0.45\textwidth]{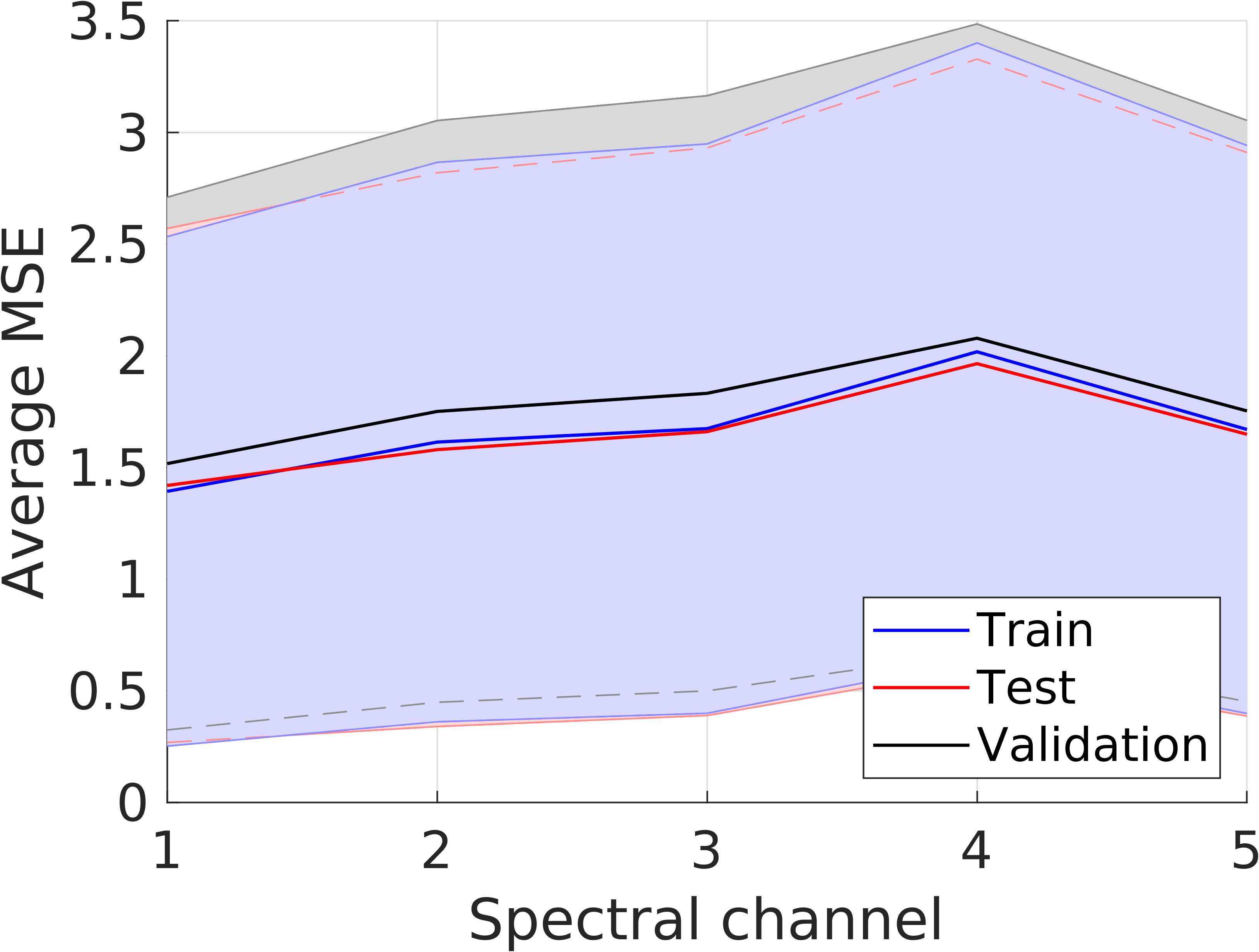}\label{Fig:spec_dist_5}} \hfill
\subfloat[]{\raisebox{2ex}{\includegraphics[width=0.5\textwidth]{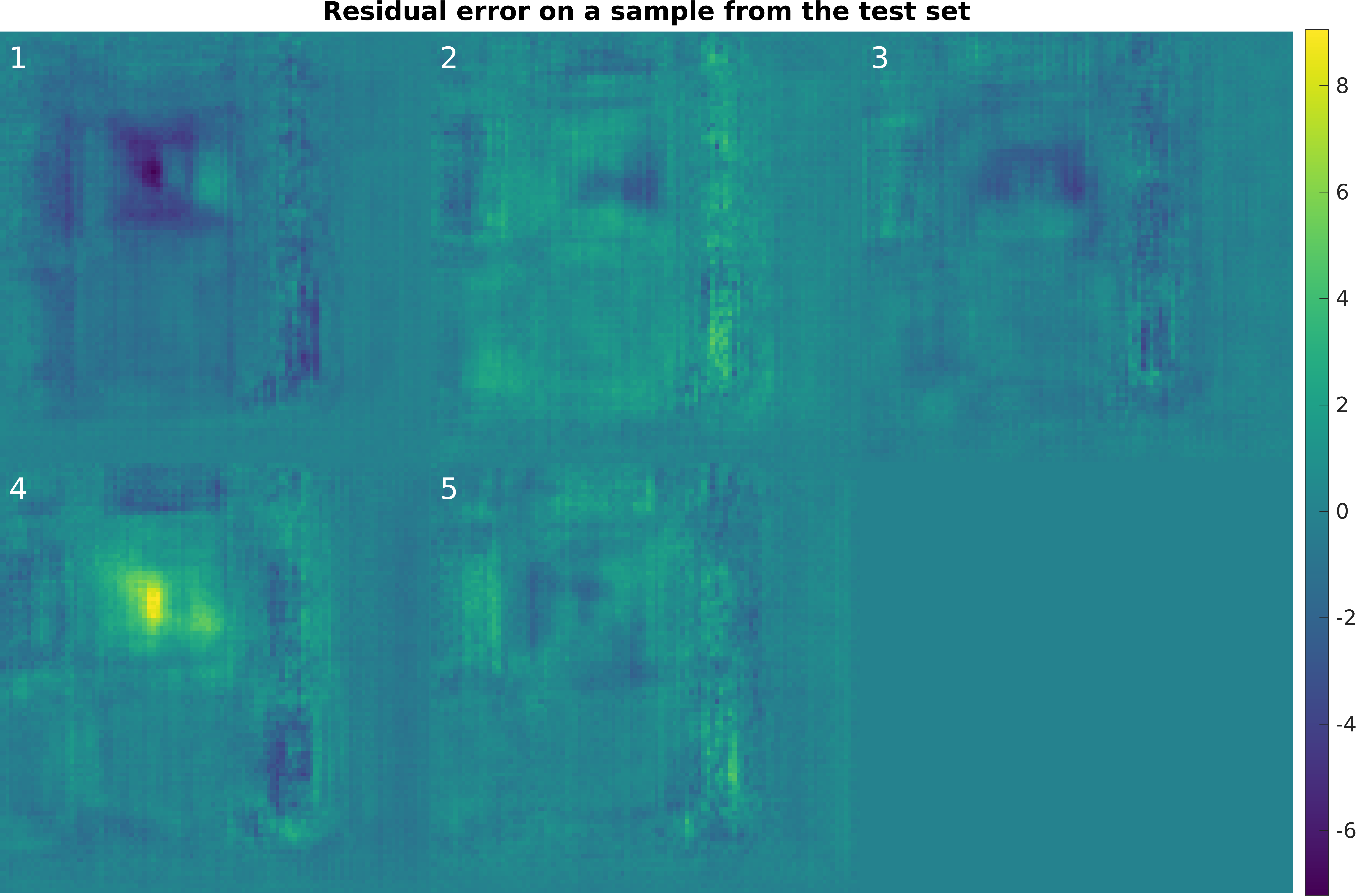}\label{Fig:spec_res_5}}}
\caption{(a) Average MSE for 5 spectral channels, where the shaded areas indicate one standard deviation. (b) The residual error from a sample in the test set for visualization of the MSE variation among the channels.
The channel index is highlighted in white.}
\end{center}
\end{figure}

We also show the average MSE for each spectral channel in Fig.~\ref{Fig:spec_dist_5} for the sparse-cropping samples\footnote{Sparse-cropping samples are chosen as they have larger MSE than the full-cropping ones.} from the training, validation and test sets, respectively. The training and validation sets behave similarly with smaller MSEs as they are comprised of the same cubes of $100 \times 100 \times 5$, whereas the training set consists of completely different cubes. 
The variation in MSE can be visualized in Fig.~\ref{Fig:spec_res_5} where the residual error of a sample from the test set is presented with the channel index indicated in the top-left corner.
The MSE fluctuates roughly between 1 and 2, which is smaller than that of the 25-channel scenario in Section~\ref{subsec:imags_25}.

\begin{figure}[htp]
\begin{center}
\includegraphics[width=1\textwidth]{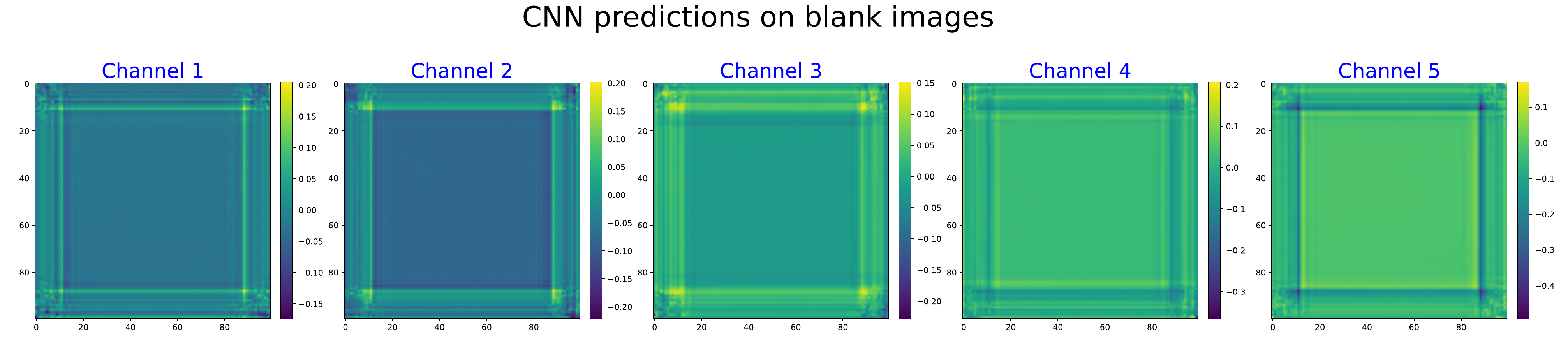}
\caption{Model predictions on the five spectral channels for blank input images. The residual errors are much smaller than the rest of the training data as shown in Table~\ref{tab:error_break}.
}
\label{Fig:blank_5}
\end{center}
\end{figure}
As mentioned in Sec.~\ref{sec:net_a}, to ensure that the network is free of bias because it only sees CTIS images of the ColorChecker, the samples of blank images are also included in the training data.
In Fig.~\ref{Fig:blank_5}, the residual errors exhibit patterns of crossing at right angles
especially around the four corners.
The magnitude of the residual errors, nonetheless, are much smaller compared to those of the ColorChecker as can be seen in Table~\ref{tab:error_break}.
As we only use convolutional layers without the pooling operation, these patterns originate from the boundary area of the input images.
To be more concrete, the network has to remove the blank area\footnote{It results from image preprocessing which cuts out the blank corners from the synthetic image as shown in Fig.~\ref{Fig:5_imags_b}.} around the boundary in the input images~(see, e.g., the bottom row of Fig.~\ref{Fig:5_imags_a}) and reconstruct each spectral channel of hyperspectral cubes, resulting in the residual crossing patterns.

\begin{figure}[htp]
\begin{center}
\includegraphics[width=0.65\textwidth]{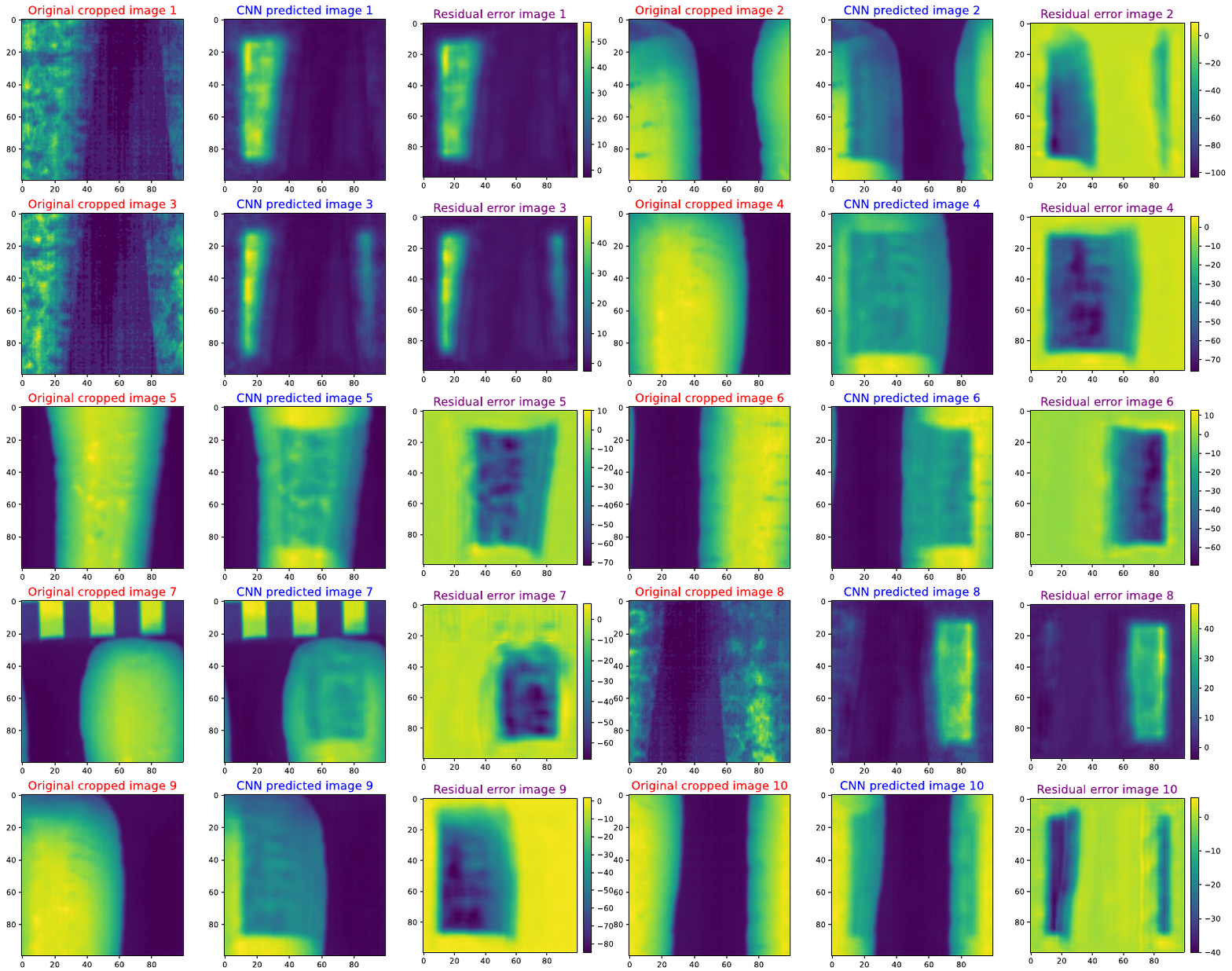}
\caption{Network predictions on wholly unseen carrot images. The network can capture the overall shapes but fall short of reconstructing details. 
}
\label{Fig:images_5_car_pre}
\end{center}
\end{figure}

Having demonstrated the network's ability to 
reconstruct the hyperspectral cube from synthetic 
CTIS images of the ColorChecker with five wavelength channels, 
we want to consider the performance on different kinds of images -- in particular images of practical relevance --- as well as a larger number of output channels.

\subsection{Predictions on carrot images}
First we simply take the network trained on the ColorChecker images and apply it directly to reconstruct the hypercube from the synthetic CTIS images of carrots~(bottom row of Fig.~\ref{Fig:5_orig}).
The reason we choose carrot images is that the corresponding hyperspectral cubes can be used to detect frost damages in carrots~\cite{8809748}.
The results are presented in Fig.~\ref{Fig:images_5_car_pre}. 
The model, overall, is able to capture the shape of carrots for the most part but fail to reproduce details precisely.
In particular noticeable residual errors manifest themselves as approximate rectangles in dark blue.
This is not surprising but can be ascribed to the bias inherited from ColorChecker images,
which are basically comprised of rectangles.

\section{Generalization to 25 spectral channels with ColorChecker and  carrot hyperspectral cubes\label{subsec:imags_25}}

\begin{figure}[htp]
\begin{center}
\includegraphics[width=0.63\textwidth]{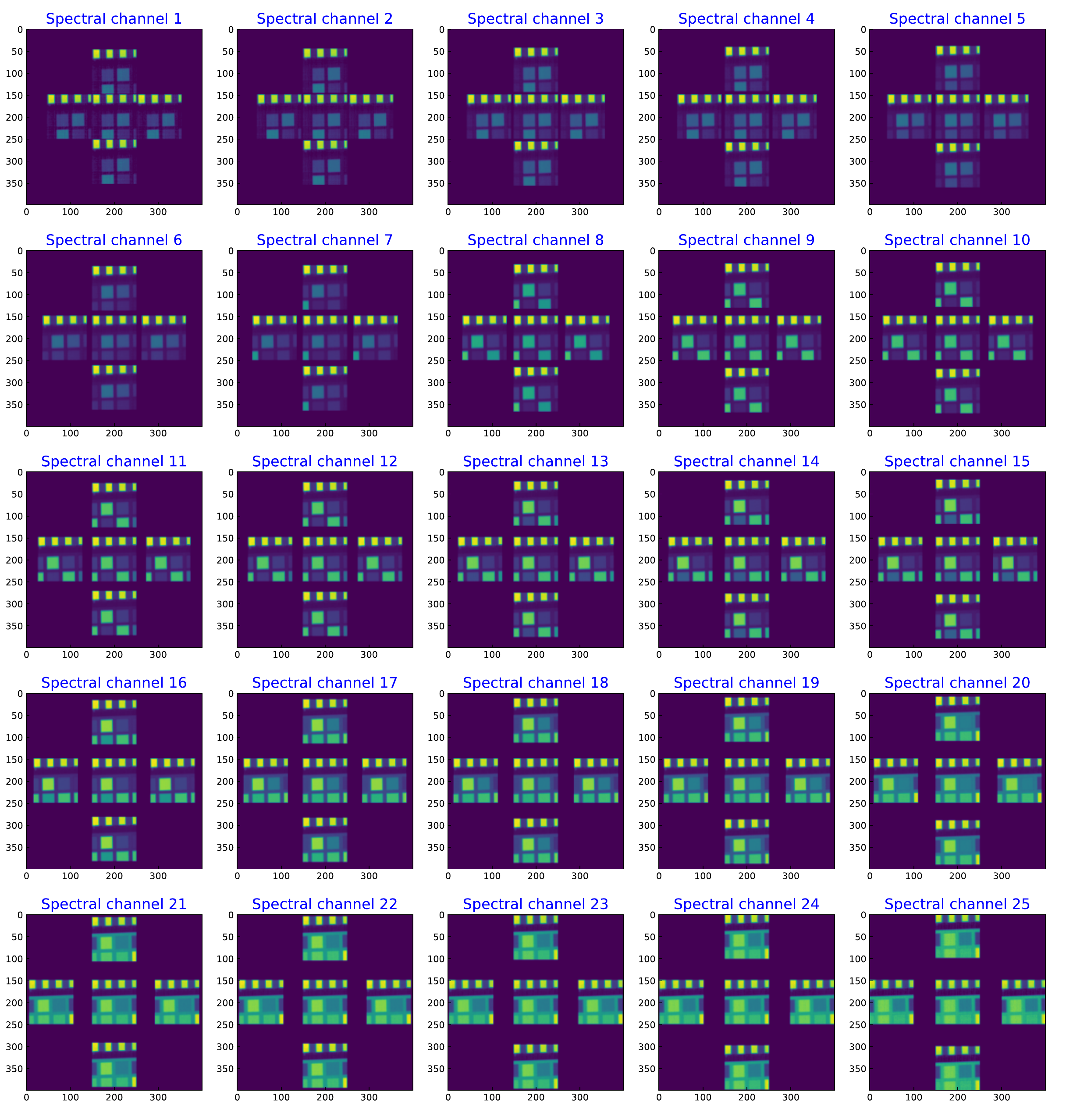}
\includegraphics[width=0.35\textwidth]{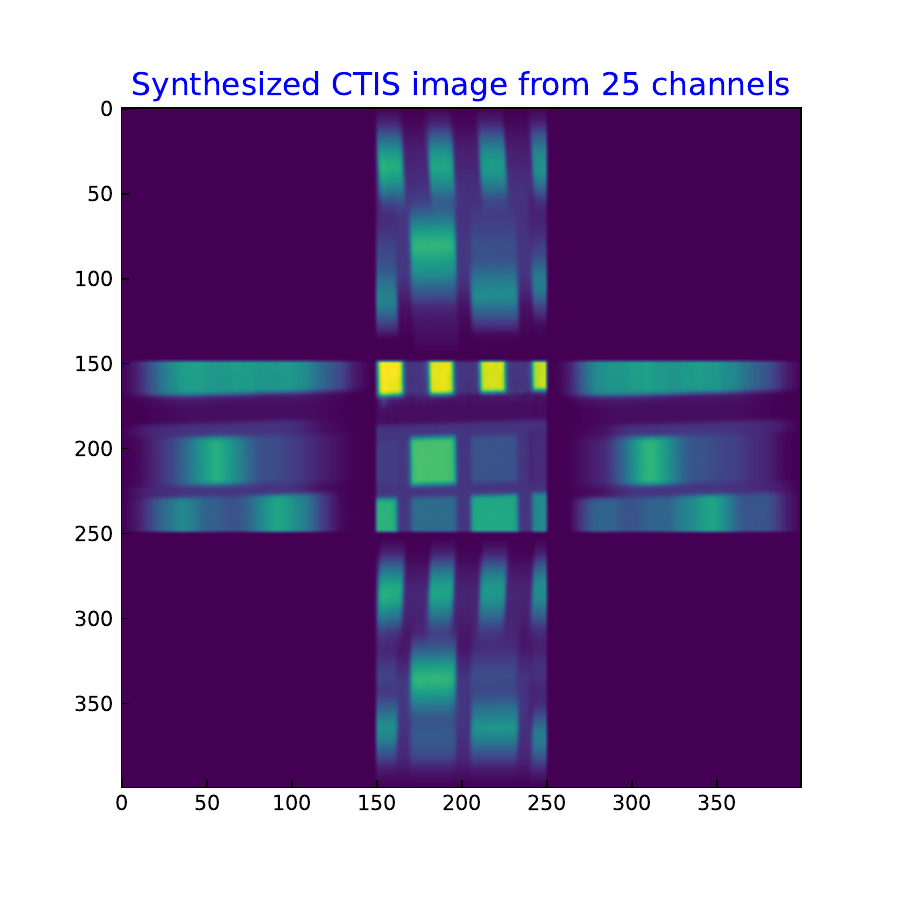} 
\caption{Left panel: Cropped, diffraction images of the 25 spectral channels. Right panel: The resulting simulated CTIS image.
}
\label{Fig:25_imags}
\end{center}
\end{figure}

Now, we are in a position to tackle the more challenging task of reconstructing hyperspectral cubes with 25 spectral
channels from synthetic CTIS images. 
We closely follow the procedure of data generation in Sec.~\ref{sec:data_CNN}.
To 
create a synthetic CTIS image, we take 25 channels from the original hyperspectral cube  -- obtained with the  aforementioned pushbroom system --  and crop this $200\times400\times25$ hyperspectral cube into $100\times100\times25$ hyperspectral cubes. For each hyperspectral cube, the 25 channels are superimposed with shifts of (2,~4,~6,~\dots,~50) pixels for the first-order diffraction
as shown in Fig.~\ref{Fig:25_imags}. The left panel shows the original channels while the right panel represents the resulting CTIS image. 
In this case, the input images become $200\times200$, nearly three times larger than the previous one.
As demonstrated in Sec.~\ref{subsec:CNN_pred}, it is more difficult for the network to make correct predictions on samples from sparse cropping than full cropping.
Therefore, we focus only on the more challenging data from sparse cropping, consisting of half a million samples.
In addition, the blank images~(0.2 million) are included into the training data as before. 

\begin{figure}[htp]
\begin{center}
\includegraphics[width=\textwidth]{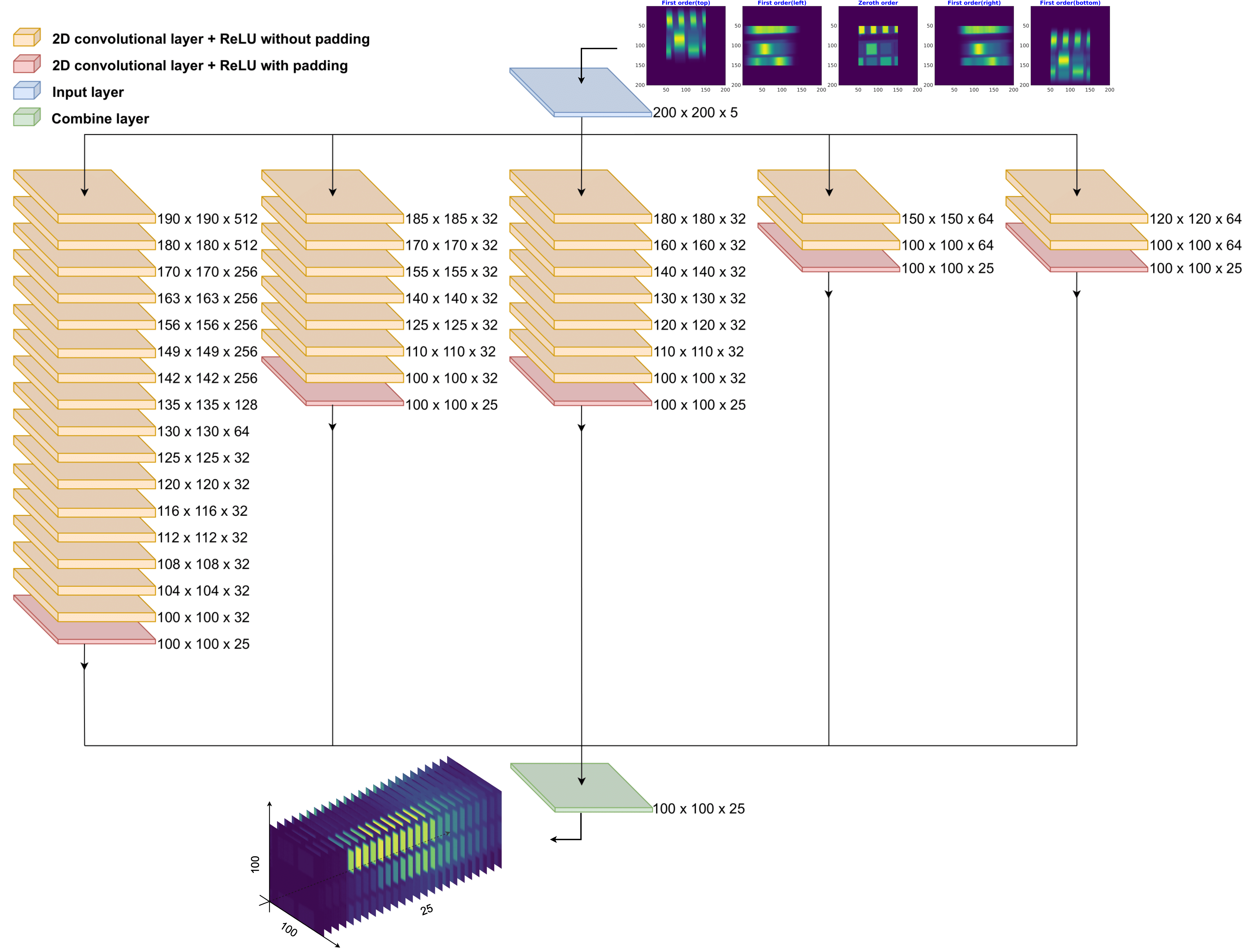}
\caption{CNN architecture for outputs of 25 spectral channels.
}
\label{Fig:CNN_25}
\end{center}
\end{figure}

\begin{figure}[htp]
	\centering
	
	\subfloat[]{\includegraphics[width=0.5\textwidth]{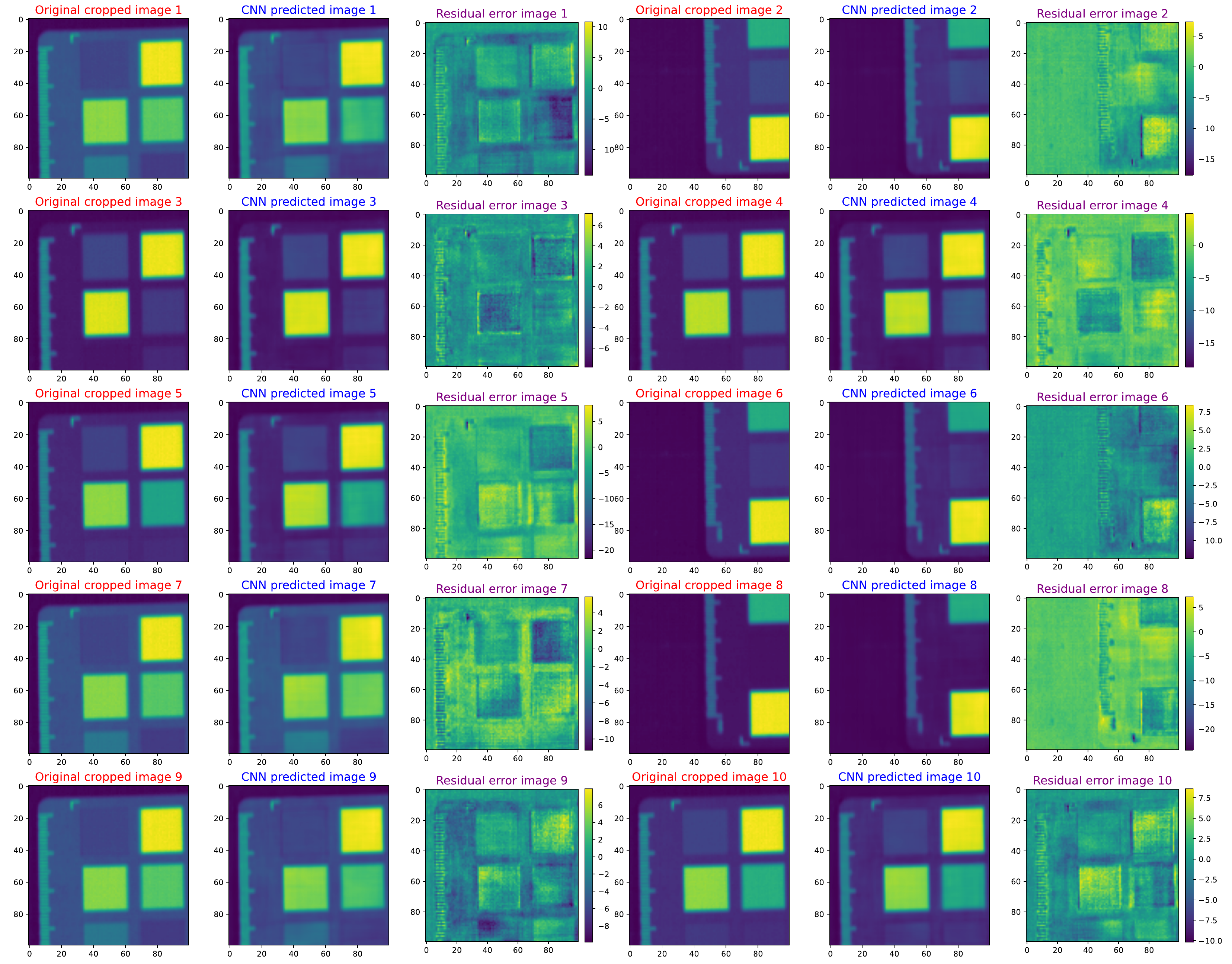}\label{Fig:images_25_pred}}
	\hfill
	\subfloat[]{\includegraphics[width=0.497\textwidth]{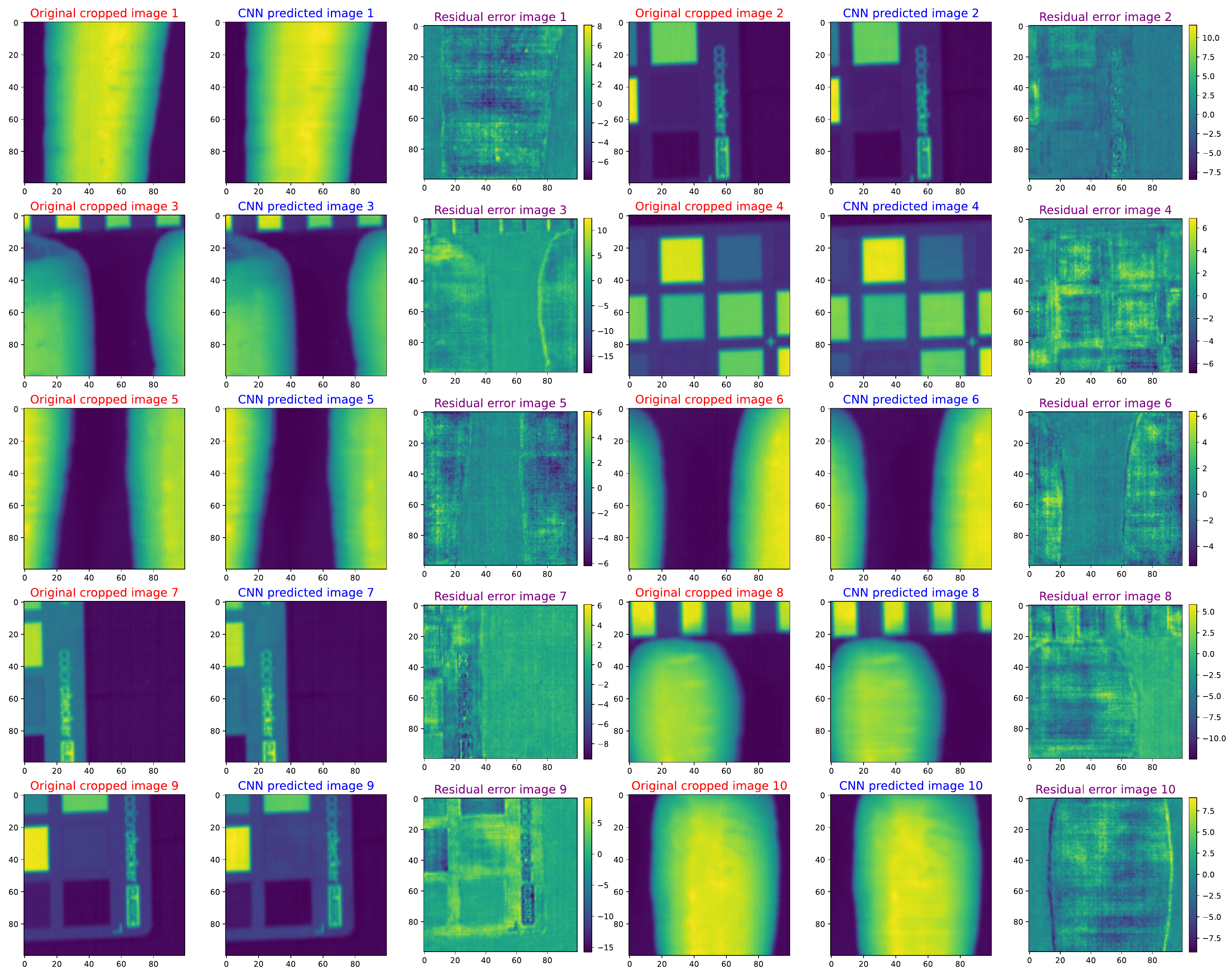}\label{Fig:images_25coca_pred}}
	\hfill
	\caption{(a)~Similar to Fig.~\ref{Fig:images_5_pred} but with 25 spectral channels. (b)~Similar to Fig.~\ref{Fig:images_25_pred} but with  the inclusion of ColorChecker and carrot cubes. The 15\textsuperscript{th} channel is shown here.}
	\label{Fig:images_25_tot}
\end{figure}

Due to the larger dimensionality~(roughly three times larger) of input images and the higher number of output channels~(five times more) for each sample, compared to the 5-channel scenario, the second network shown in Fig.~\ref{Fig:CNN_25} becomes approximately twelve times bigger than before -- 85.9 million trainable parameters in total 
versus 6.79 million for the first network.
The comparison, similar to Fig.~\ref{Fig:images_5_pred}, between the true images and model predictions is displayed
in Fig.~\ref{Fig:images_25_pred} for the training~(three columns on the left) and test~(three on right) data. Clearly, the model also generalizes satisfactorily to completely unseen data with quite similar accuracy.

Finally, we include into the training data another 0.5 million samples of carrot images.
The results are illustrated in Fig.~\ref{Fig:images_25coca_pred} -- the network is able to simultaneously deal with the different types of images very well
with the decent ability of generalization.
The model performance is summarized in Table~\ref{tab:error_break_25} which breaks down into two types of training data
-- the ColorChecker and the ColorChecker plus carrots. It is evident that the model can reproduce two different types of hyperspectral cubes but performs
slightly better on the carrot images. Note that in light of the complexity of the model and 
hence a very time-consuming training process
\begin{figure}[htp]
\begin{center}
\subfloat[]{\raisebox{11ex}{\includegraphics[width=0.45\textwidth ]{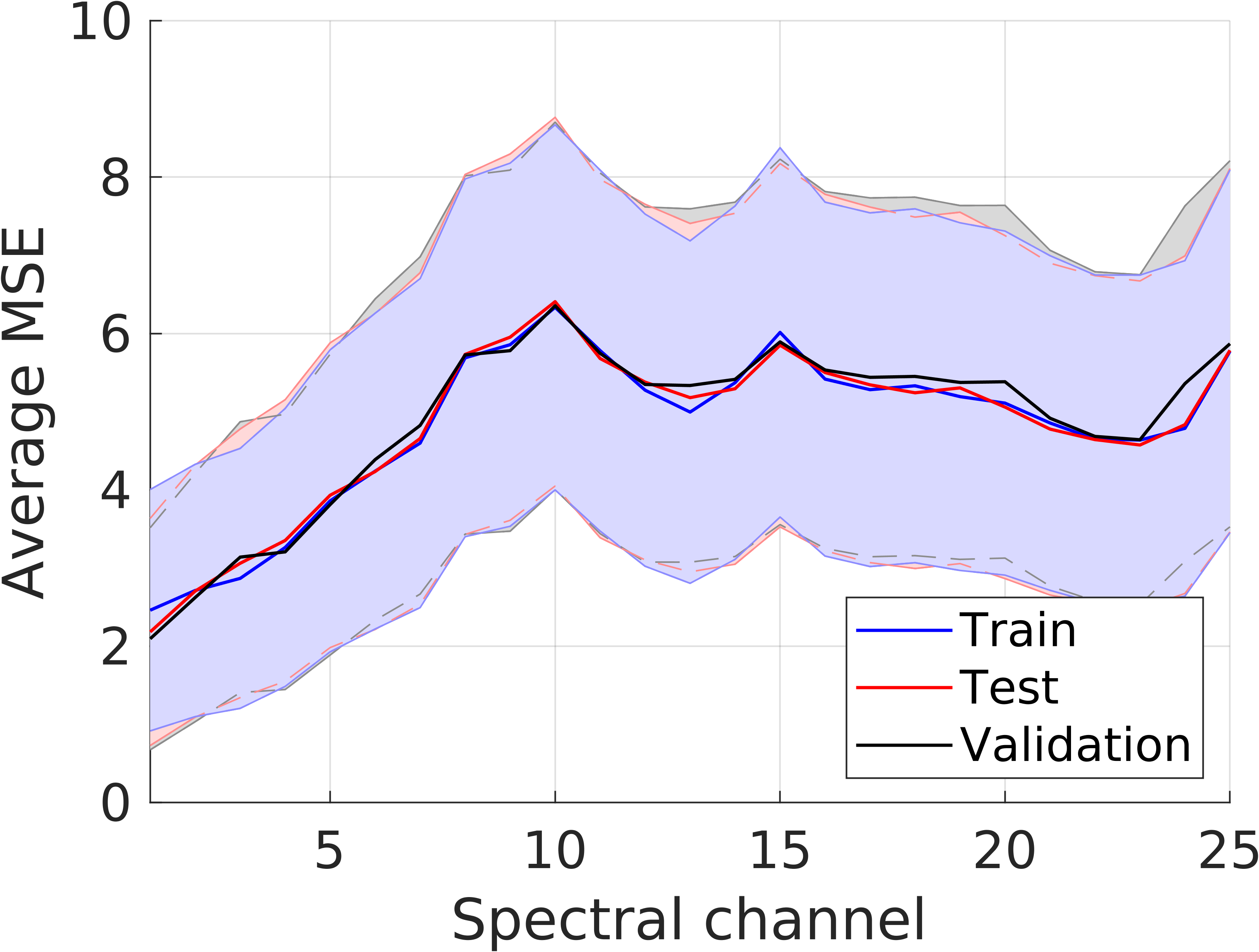}\label{Fig:spec_dist_25}}}
\hfill
\subfloat[]{\includegraphics[width=0.5\textwidth]{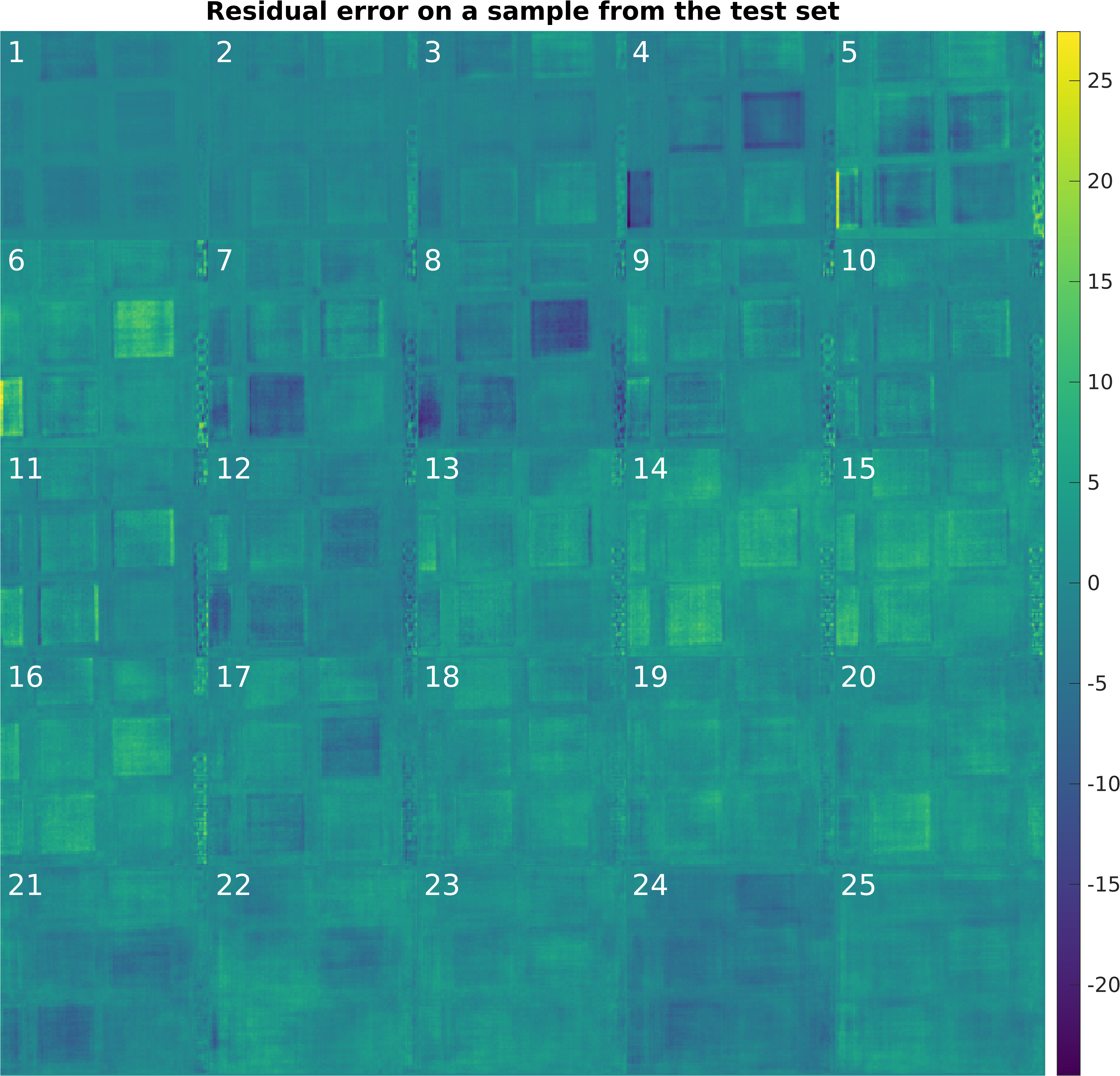}\label{Fig:spec_res_25}}
\caption{Average MSE for 25 spectral channels, where the shaded areas indicate one standard deviation.  (b) The residual error from a sample in the test set for visualization of the MSE variation among the channels. The channel index is highlighted in white.}
\end{center}
\end{figure}
we train the model only for 25 epochs
in both cases as opposed to 40 epochs in the case of 5 spectral channels.
It is expected that the performance will increase with more epochs and might potentially match the previous results of 5-channel outputs with MSE $\sim 1$.
Again, we also show the average MSE of each spectral channel in Fig.~\ref{Fig:spec_dist_25} for the training, validation and test sets, respectively.
The residual error of a sample from the test set is shown in Fig.~\ref{Fig:spec_res_25} to display the fluctuation in MSE (a comparison with EM for the same sample is shown in Appendix \ref{sec:appSpectral}), which is more pronounced than the 5-channel case in Fig.~\ref{Fig:spec_res_5}.
The residual error images, $100 \times 100$ for each channel, are concatenated into a $500 \times 500$ image with the channel index in the top-left corner.

It is worth mentioning that a forward pass through the network (making a prediction) takes only 13~ms on average. It implies the network is capable of carrying out
the reconstruction of 3-D data cubes {\it on the fly} while a CTIS camera is capturing images. In this way, the subsequent data analysis based on the reconstructed cubes can also be executed in real-time and provide immediate feedback on the system; for instance, on how the deployment of the CTIS camera or the imaged object should be adjusted.

\begin{table}[htp]
	\centering
\begin{tabular}{P{2.4cm}<{\raggedright}P{2.4cm}P{2.5cm}<{\raggedright}P{2.5cm}P{1.9cm}P{1.9cm}P{1.9cm}}
	\toprule
	Training set & \# of sample & Test samples & \# of samples & MSE & MAE & PSNR \\ \midrule
	\multirow{2}{=}{$ C_{\text{train}}$ + blank image} & \ $0.7\cdot 10^6$ & $C_{\text{train}}$ &  $0.5\cdot 10^6$ & 5.32 & 1.56 & 40.9 \\ \cmidrule{2-7}
	 & \ $0.7\cdot 10^6$ & $C_{\text{test}}$ & $0.5\cdot10^6$ & 5.69 & 1.59 & 40.6 \\ \hline
	\multirow{4}{=}{$C_{\text{train}}$ + $Car_{\text{train}}$+ blank images} & \multirow{2}{*}{ $1.2\cdot 10^6 $} & \multirow{2}{2.8cm}{$C_{\text{train}}+Car_{\text{train}}$ $ (C_{\text{train}}, Car_{\text{train}})$} & \multirow{2}{*}{\ \ $1\cdot 10^6 $} & \multirow{2}{1.9cm}{\centering 4.82 (5.33,~4.52)}& \multirow{2}{1.9cm}{\centering 1.45 (1.55,~1.37)} & \multirow{2}{1.9cm}{\centering 41.3 \\(40.9,~41.6)} \\ \\ \cmidrule{2-7}
	 & \multirow{2}{*}{ $1.2\cdot 10^6 $} & \multirow{2}{2.8cm}{$C_{\text{test}}+Car_{\text{test}}$ $ (C_{\text{test}}, Car_{\text{test}})$} & \multirow{2}{*}{\ \   $1\cdot10^6$} & \multirow{2}{1.9cm}{\centering 5.03 \\ (5.59,~4.28)} & \multirow{2}{1.9cm}{\centering 1.48 (1.58,~1.35)} & \multirow{2}{1.9cm}{\centering\centering 41.1 (40.7,~41.3)} \\ \\
	\bottomrule
\end{tabular}
	\caption{Similar to Table~\ref{tab:error_break} but for 25 spectral channels. The ColorChecker is abbreviated as $C_{\text{dataset}}$ and carrot images as $Car_{\text{dataset}}$. The network is separately trained with two data sets, ColorChecker only and ColorChecker plus carrot images.
	For the latter case, we present the errors for images of the ColorChecker and carrots combined as well as individual ones~(values in parentheses).}
	\label{tab:error_break_25}
\end{table}

\section{CTIS expectation maximization reconstruction}

In this Section, we compare the CNN approach with the standard expectation maximization~(EM) algorithm~\cite{shepp_maximum_1982}, utilized in the CTIS reconstruction.
We begin with the creation of a system matrix followed by the cube reconstruction using the EM algorithm.
We shall see that the CNN method has a better performance in reproducing hyperspectral cubes than the EM algorithm, and has a shorter process time when it comes to cube reconstruction.

\subsection{System matrix generation and EM}
To investigate the performance of the network relative to the standard reconstruction approach, an equivalent CTIS simulation is performed. The simulation is based on the standard approach to reconstruct a hyperspectral cube from an acquired CTIS image, which assumes that the system is described by the linear imaging
equation~\cite{hagen_fourier_2007}:
\begin{align}\label{Eq:ImagEq}
	\boldsymbol{g} = \boldsymbol{H}\boldsymbol{f}
\end{align}
where $\boldsymbol{g}$ is the vectorized $q\times q$ CTIS image\footnote{Namely, $\boldsymbol{g}$ is a column vector with $q^2$ elements.} with $q=3x+4b$, $\boldsymbol{H}$ is the $q^2\times r$ system matrix and $\boldsymbol{f}$ is the vectorized hyperspectral cube with $r=x \cdot y \cdot b$ voxels, where $x,y$ and $b$ denote the two spatial dimensions and spectral channels, respectively. The system matrix $\boldsymbol{H}$ maps the sensitivity of the $i$-th voxel in $\boldsymbol{f}$ to the $j$-th pixels in $\boldsymbol{g}$ -- corresponding to the five projections shown in Fig.~\ref{Fig:5_imags_b}. 
The system matrix $\boldsymbol{H}$ is constructed by assuming spatial shift-invariance and an ideal CTIS, i.e. each voxel in $\boldsymbol{f}$ is projected once into each diffraction order. Spatial shift-invariance assumes that shifting a voxel by a given distance within the hyperspectral cube yields a CTIS image in which the corresponding pixels in the zero- and first diffraction orders are shifted by the same distance in the same direction.
Thus, the $i$-th voxel, $f_i$, is projected onto a $q\times q$ CTIS image, which is vectorized and arranged as columns in $\boldsymbol{H}$. This process is repeated for all voxels in $\boldsymbol{f}$ until $\boldsymbol{H}$ has been completely constructed. Thus, each column in $\boldsymbol{H}$ only has five nonzero elements corresponding to the five projections, which leads to the natural implementation of $\boldsymbol{H}$ as a sparse matrix since its sparsity is $(q^2-\text{nonzero elements})/q^2 = (400^2-5)/400^2 = 0.99997$. 

To effectively reconstruct the hyperspectral cube $\boldsymbol{f}$ from an acquired CTIS image $\boldsymbol{g}$  requires solving Eq.~\eqref{Eq:ImagEq} through the inversion of the system matrix $\boldsymbol{H}$. However, since $\boldsymbol{H}$ is noninvertible, the iterative EM algorithm is utilized to obtain $\boldsymbol{f}$ instead.
The EM algorithm consists of an expectation and a maximization step. First, an estimated CTIS image $\hat{\boldsymbol{g}} = \boldsymbol{H} \hat{\boldsymbol{f}}^{(k)}$ is computed in the expectation step. In the subsequent maximization step, a correction factor for every voxel in the estimated hyperspectral cube $\hat{\boldsymbol{f}}^{(k)}$ is computed as a back-projection of the ratio of the captured and estimated CTIS image and normalized by the summed rows of $\boldsymbol{H}$.
All in all, we have:
\begin{align}\label{eq:EM}
	\hat{\boldsymbol{f}}^{(k+1)}	 = \frac{\hat{\boldsymbol{f}}^{(k)}}{\sum_{i=1}^{q^2} H_{ij}}\odot \left( \boldsymbol{H}^T \frac{\boldsymbol{g}}{\boldsymbol{H}\hat{\boldsymbol{f}}^{(k)}}\right)
\end{align}
where $k$ is the iteration index, $\hat{\boldsymbol{f}}^{(k)}$ is the $k$-th estimate of the hyperspectral cube, $\sum_{i=1}^{q^2} H_{ij}$ is the vectorized summation of rows in $\boldsymbol{H}$, $\boldsymbol{H}^T$ is the transposed system matrix and $\odot$ denotes the Hadamard or element-wise product. Notice, that Eq.~\eqref{eq:EM} combines the  expectation and maximization steps into a single step.

The algorithm is typically initialized with either $\hat{\boldsymbol{f}}^{(0)} = ones(r,1)$~\cite{wilson_reconstructions_1997} or $\hat{\boldsymbol{f}}^{(0)} = \boldsymbol{H}^T\boldsymbol{g}$~\cite{descour_computed-tomography_1995}. The latter is assumed in this work. As 10-30 EM iterations are typically required~\cite{white_accelerating_2020}, we carry out 20 iterations. Both the construction of $\boldsymbol{H}$ and reconstruction of $\boldsymbol{f}$ are implemented in \texttt{MATLAB} with
the help of built-in sparse matrix manipulations. 

\subsection{Sparse EM predictions\label{subsec:CTIS_25}}
\begin{figure}[htp]
	\centering
	
	\subfloat[]{\includegraphics[width=0.4\textwidth]{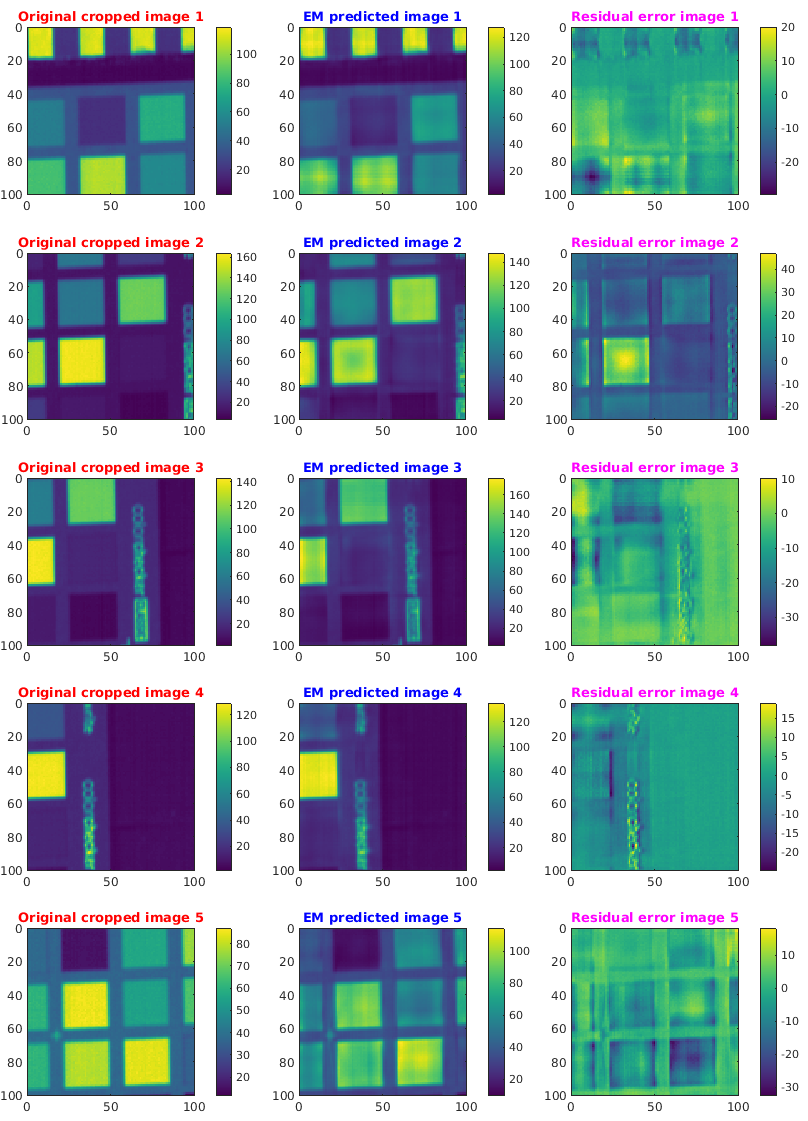}\label{Fig:25_imags_ctis_a}}
	\hfill
	\subfloat[]{\includegraphics[width=0.4\textwidth]{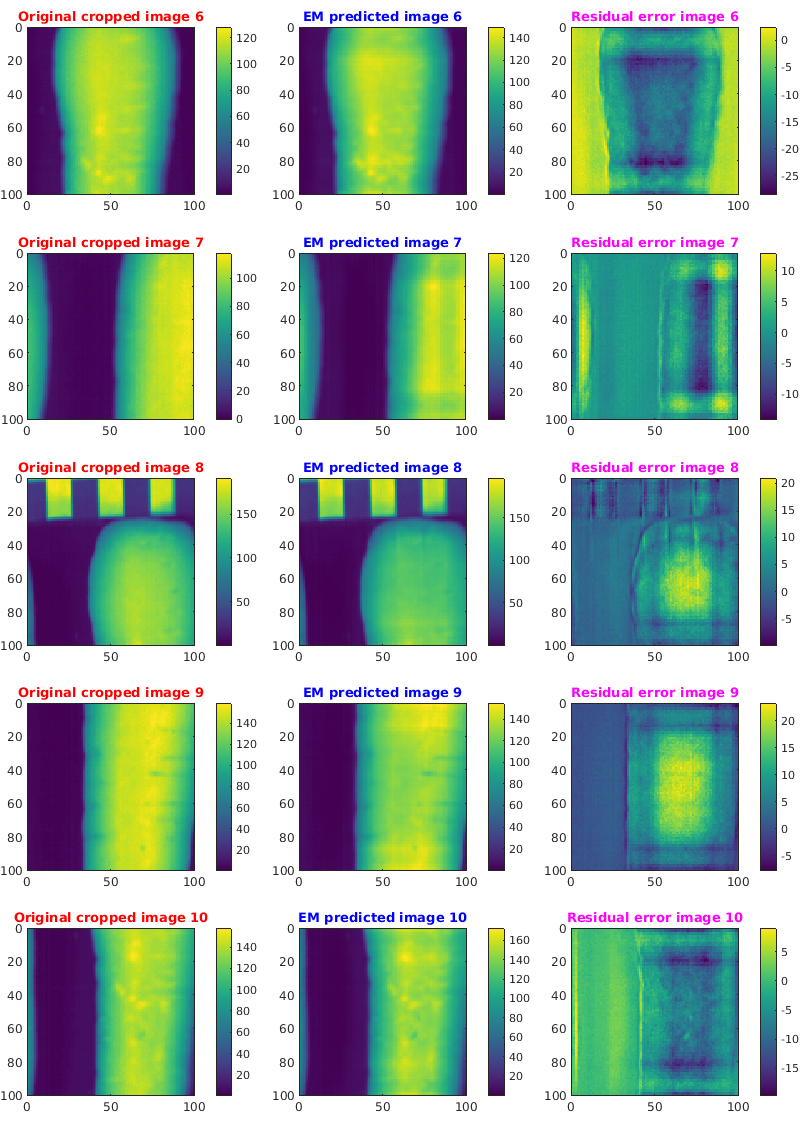}\label{Fig:25_imags_ctis_b}}
	\hfill
	\caption{Comparison of original (left column), reconstructed (central column) and residual (right column) spectral channel 15 from $100\times100\times25$ hyperspectral cubes for (a)  the ColorChecker and (b) carrots based on the sparse EM reconstruction algorithm.}
	\label{Fig:25_imags_ctis}
\end{figure}

The reconstruction algorithm requires no training data, and therefore, both the training and unseen data are reconstructed. From the data, only 2000 CTIS images have been chosen since the EM algorithm is run on a regular laptop equipped with an 11th Gen Intel\textregistered \ Core\texttrademark \  i7-1165G7 2.80 GHz CPU. The original and predicted channels as well as the residual error for the ColorChecker and carrots are shown in Fig.~\ref{Fig:25_imags_ctis_a} and Fig.~\ref{Fig:25_imags_ctis_b}, respectively. Similar rectangular reconstruction artifacts as seen in Fig.~\ref{Fig:images_25_pred} and Fig.~\ref{Fig:images_25coca_pred} are also visible for the EM reconstructed channels. The average computation time for 20 EM iterations over the 2000 hyperspectral cubes is measured to be $94.7 \pm 3.5\text{~ms}$, where the uncertainty indicates the standard deviation of the measured computation times. 
This is comparable~(or even faster if taking into account hardware and implementation differences) to the current state-of-the-art algorithm, which exploits spatial shift-invariance,
by White et al.~\cite{white_accelerating_2020}. It reconstructed a smaller $89\times80\times24$ cube in 0.5 and $1.2\text{~s}$ for 10 and 25 EM iterations, respectively.
However, the EM algorithm is still slower than a forward pass in the CNN which takes 13~ms on average.
We should point out that it is not a completely fair comparison since first the network  needs the time-consuming training stage\footnote{It takes around three days to train the first CNN with five spectral channels and nearly two weeks for the second CNN of 25 spectral channels.} before making decent predictions and second
the EM algorithm is carried out on the regular laptop as opposed to the four powerful GPUs in the CNN case.
However as a proof of principle study, these results demonstrate the potential of the CNN approach on deciphering CTIS images on the fly.

\begin{figure}[htp]
	\begin{center}

			\subfloat[]{\includegraphics[width=0.48\textwidth]{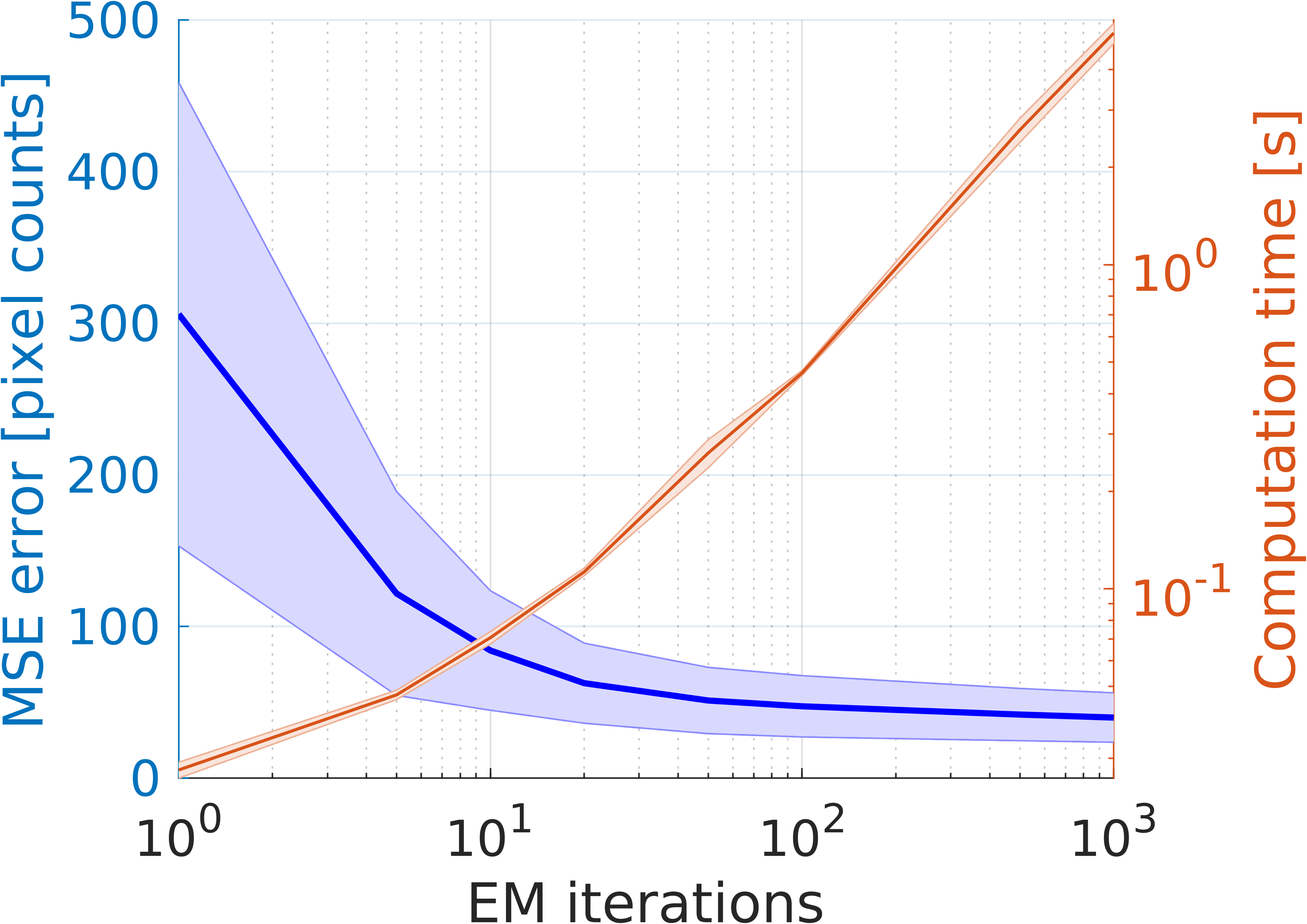}\label{Fig:ErrorvsEMiter_a}}
			\hfill
			\subfloat[]{\includegraphics[width=0.45\textwidth]{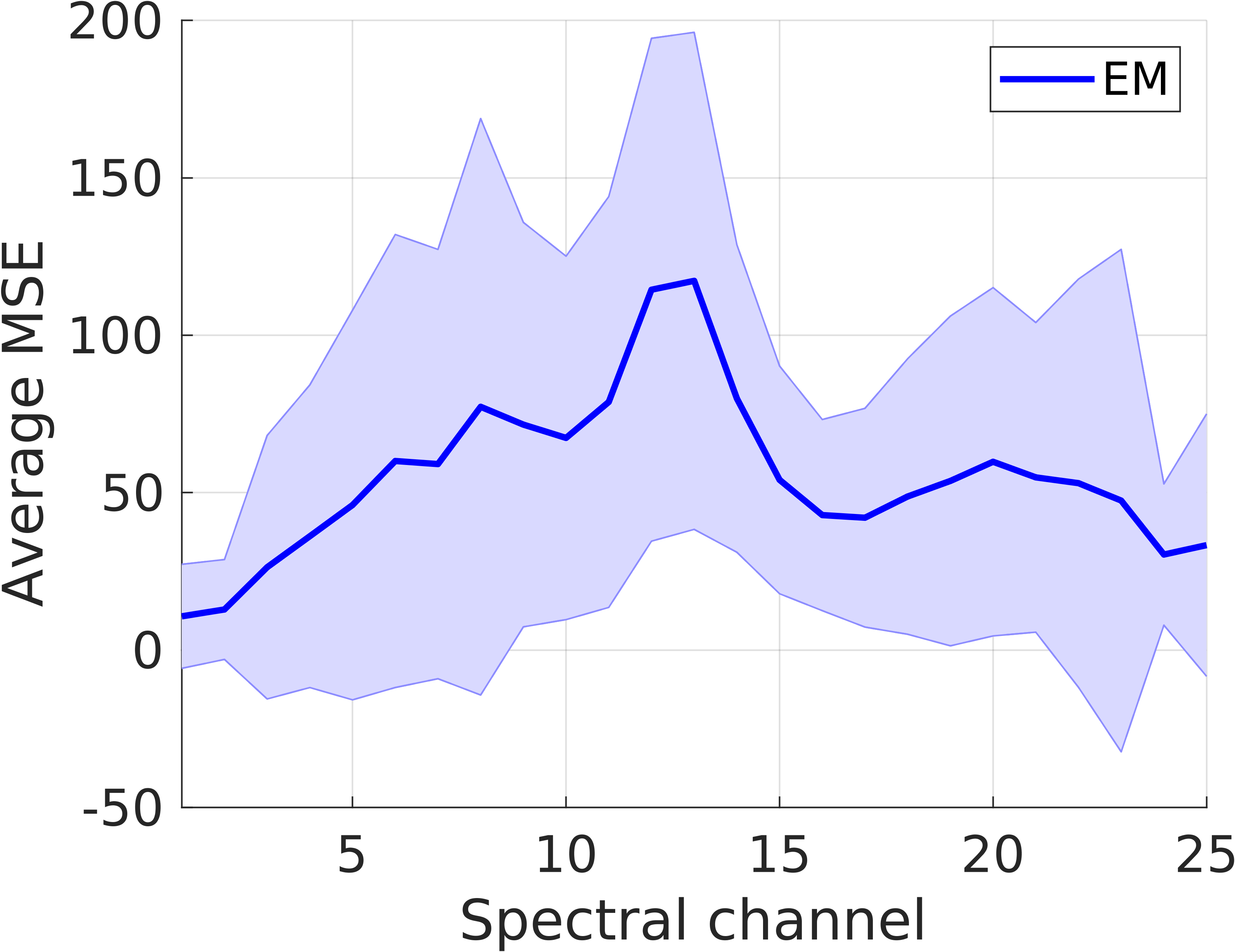}\label{Fig:ErrorvsEMiter_b}}
		\caption{(a) The MSE and computation time for the EM algorithm as a function of EM iterations. The computed values and shaded areas correspond to the mean and standard deviation of 20 randomly chosen hyperspectral cubes, respectively. (b) The average MSE for 25 spectral channels for EM, where the values are calculated based on 2000 CTIS images.
		Again, the EM algorithm is executed on the regular laptop. }
		\label{Fig:ErrorvsEMiter}
	\end{center}
\end{figure}

The average MSE, MAE and PSNR of the EM algorithm are summarized in Table~\ref{tab:error_ctis_25}: For the ColorChecker, we obtain MSE, MAE and PSNR of 55.50, 4.62 and 30.7, respectively, which are significantly higher than those of the networks in Table~\ref{tab:error_break_25}. The MSE, MAE and PSNR for the carrots are computed to be 130.64, 6.62 and 27.0, respectively. Fig.~\ref{Fig:ErrorvsEMiter_b} shows the average MSE for each spectral channel for the 2000 reconstructed hyperspectral cubes using EM.
\begin{table}[htp]
 	\centering
	\begin{tabular}{P{7cm}<{\raggedright}  P{3cm} P{3cm} P{3cm} }
		\toprule
		& MSE & MAE & PSNR\\
		\midrule
		ColorChecker & 55.50 &  4.62 & 30.7 \\
		Carrots  & 130.64  &  6.62 & 27.0 \\ 	
		\bottomrule
	\end{tabular}
	\caption{The mean MSE, MAE and PSNR for 2000  reconstructed $100\times100\times25$ hyperspectral cubes of the ColorChecker and carrots using the EM algorithm with 20 iterations.}
	\label{tab:error_ctis_25}
\end{table}
The MSE for all channels is significantly larger than that of CNN, shown in Fig.~\ref{Fig:spec_dist_25}.
The significantly larger errors for carrots relative to the ColorChecker are due to the higher degree of spatiospectral multiplexing for the carrots. In other words, there is less spatial separation between regions of different spectral signatures compared to the  ColorChecker with separated square regions which eases the reconstruction~\cite{hagen_maximizing_2006, hagen_analysis_2008}. Surprisingly, this issue does not happen to the CNNs. 

It should be noted, that both the MSE and MAE decrease asymptotically while the computation time increases linearly as the number of EM iterations increases: Fig.~\ref{Fig:ErrorvsEMiter_a} shows the MSE and computation time for 1, 5, 10, 20, 50, 100, 500 and 1000 EM iterations. The computed values and error bars correspond to the mean and standard deviation of 20 randomly chosen hyperspectral cubes, respectively. The computation times are shown in a log-log plot since one and 1000 EM iterations take 23~ms and 3.92~s, respectively.
In this work, 20 iterations have been assumed, abiding by  the practice despite better performance with more iterations -- going from 20 to 1000 iterations, the MSE decreases by $36~\%$ but the computation time unfortunately increases by a factor of $\sim42$.

\begin{figure}[htp]
	\centering
	
	\subfloat[]{\includegraphics[width=0.4\textwidth]{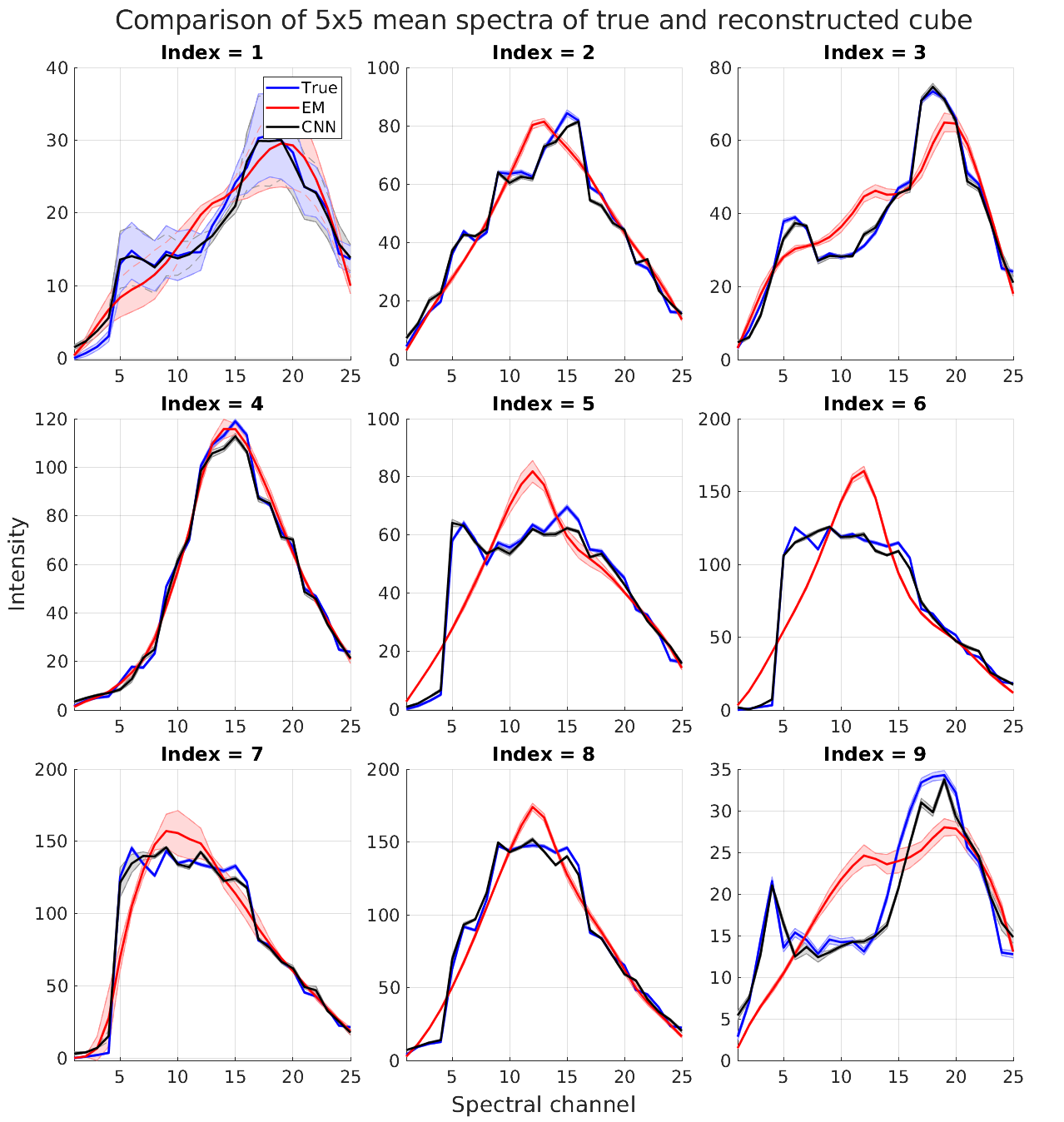}\label{Fig:specDist_check_a}}
	\hfill
	\subfloat[]{\includegraphics[width=0.4\textwidth]{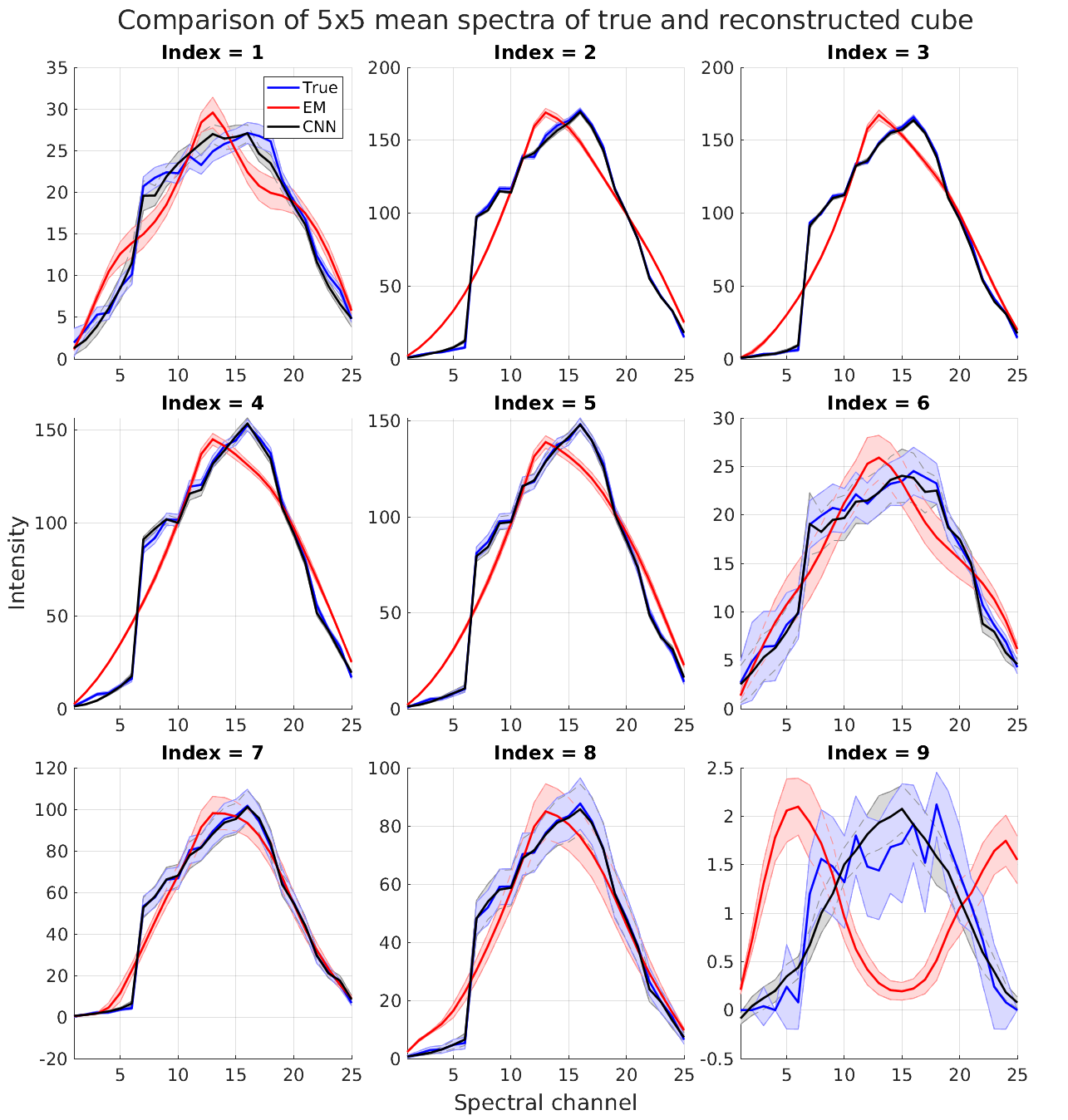}\label{Fig:specDist_check_b}}
	\hfill
	\subfloat[]{\includegraphics[width=0.19\textwidth]{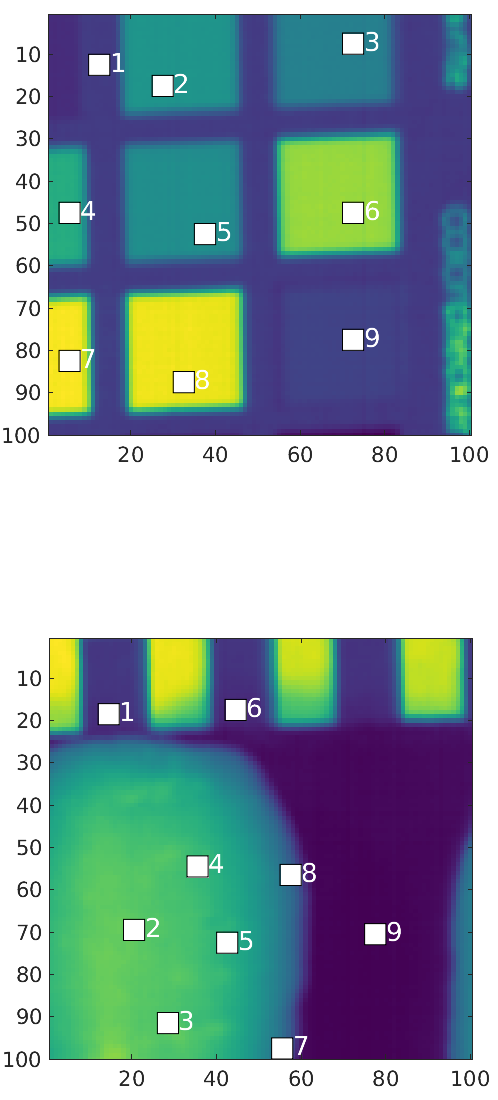}\label{Fig:specDist_check_c}}
	\hfill
	\caption{Comparison of the mean spectra of $5\times5$ pixels between the EM and CNN with respect to the true hyperspectral image of (a) the Colorchecker and (b) carrots. The position and size of the mean spectra is indicated by the white squares in (c) with the corresponding index for the ColorChecker and carrots. The shaded areas of the curve indicate one standard deviation.}
	\label{Fig:specDist_check}
\end{figure}

In Fig.~\ref{Fig:specDist_check}, a comparison of the reconstructed spectra by the EM and CNN versus the true spectra is illustrated. Fig.~\ref{Fig:specDist_check_a} and Fig.~\ref{Fig:specDist_check_b} show the mean spectra of $5\times5$ pixels squares, indicated as white squares with the corresponding index in Fig.~\ref{Fig:specDist_check_c}, for the ColorChecker and carrots, respectively. 
For all spectra the CNN significantly outperforms EM and reconstructs most of the features of the true spectra. In contrast, the EM reconstructs the overall shape of the spectra, but struggles with high-frequency components and reconstructing small features of the true spectra. This is especially evident for the square of index 9 in the carrot image, where the CNN approximately reproduces the background (noise floor) while the EM fails to capture the shape.

\section{Conclusions}
\label{sec:conclusion}

The CTIS~\cite{Okamoto:91,Th,descour_computed-tomography_1995} is a non-scanning snapshot hyperspectral imaging system. A CTIS image contains a central zeroth-order  undiffracted image surrounded by four first-order diffracted images.
Different algorithms~\cite{vose_heuristic_2007,hagen_fourier_2007,white_accelerating_2020} are used to convert multiplexed 2-D CTIS images into 3-D hyperspectral cubes, which are easier to visualize and analyze.
A CTIS camera is, however,  portable and  cheaper than hyperspectral imagers such as pushbroom
and has a wider range of applications.
However it is necessary to develop fast and reliable image reconstruction algorithms as existing ones are often prone to  long reconstruction time and mediocre precision for hyperspectral cubes  with a large number of spectral channels.

In this study, we propose a novel reconstruction 
method based on CNNs 
where the networks consist of 
five branches of convolutional layers, which allow such fast and reliable
predictions.
This method imitates decision tree ensembles, where a group of weak learners work together to form a strong learner. 
As a proof of concept, we start with a simple case where the network is required to decipher CTIS images of the ColorChecker
with five spectral channels and pixel values range from 0 to 255.
The average errors of MSE~(MAE) summarized in Table~\ref{tab:error_break} are around~(below) one for both the training and test datasets.
 It should be noted that the test data is created by completely distinct hyperspectral cubes from those of the training data. 
These results demonstrate that the network can deliver satisfactory performance and  generalize well to new (unseen) data of same basic type. 
CTIS images of carrots are then included in the training sets to demonstrate that the network can simultaneously manage different kinds of images and in particular images of direct relevance to industrial applications of CTIS imaging,
such as frost damage detection in vegetables.

However, some of the relevant industrial applications of CTIS  will require more than five spectral channels and we therefore studied the more challenging scenario with 25 spectral channels. Decent levels of precision, collected in Table~\ref{tab:error_break_25}, with MSE~(MAE) $\sim 5~(1.6)$ are obtained and the forward pass time~(the time it takes to make a prediction) takes only 13~ms. It underscores the potential of applying CNN to real-time reconstruction of hyperspectral cubes with simultaneously captured CTIS images.
Note that it is also relevant to compare CNN reconstruction time for CTIS images with those of other non-scanning snapshot HSI systems such as CASSI~\cite{gehm_single-shot_2007} and HMVIS~\cite{cao_high_2011}.
The detailed, systematic comparison on reconstruction time will be pursued in future work.
Finally, we have shown for one of the standard reconstruction methods, the EM algorithm, that the corresponding residual errors  are much larger~(by more than a factor of 10; see
Table~\ref{tab:error_ctis_25}) and the reconstruction time is also longer,
depending on the number of EM iterations and computational power available.

To summarize, we have  
demonstrated that CNNs can be used to convert CTIS images into a hyperspectral cube in an efficient and precise manner.
Our work lays the foundation for future studies on network architecture and optimization with the goal of realizing fully automated and versatile CNNs for reconstruction of hyperspectral cubes on the fly which will warrant a broad range of applications
where real-time 2-D spectroscopy is in demand.
\acknowledgments

MTF and WCH acknowledge partial funding from the Independent Research Fund Denmark, grant number 
DFF 6108-00623, The Villum Foundation and a CenSec grant funded by The Danish Ministry of Higher Education and Science (CenSec). MSP acknowledge partial funding from the Innovation Fund Denmark (IFD) under File No. 1044-00053B.
This work was performed using the \href{https://escience.sdu.dk/index.php/ucloud/}{UCloud} computing and storage resources, managed and supported by eScience center at SDU.

\appendix 
\section{Experimenting with different network architectures}
\label{sec:app}
\begin{figure}[htp]
\begin{center}
\includegraphics[width=0.95\textwidth]{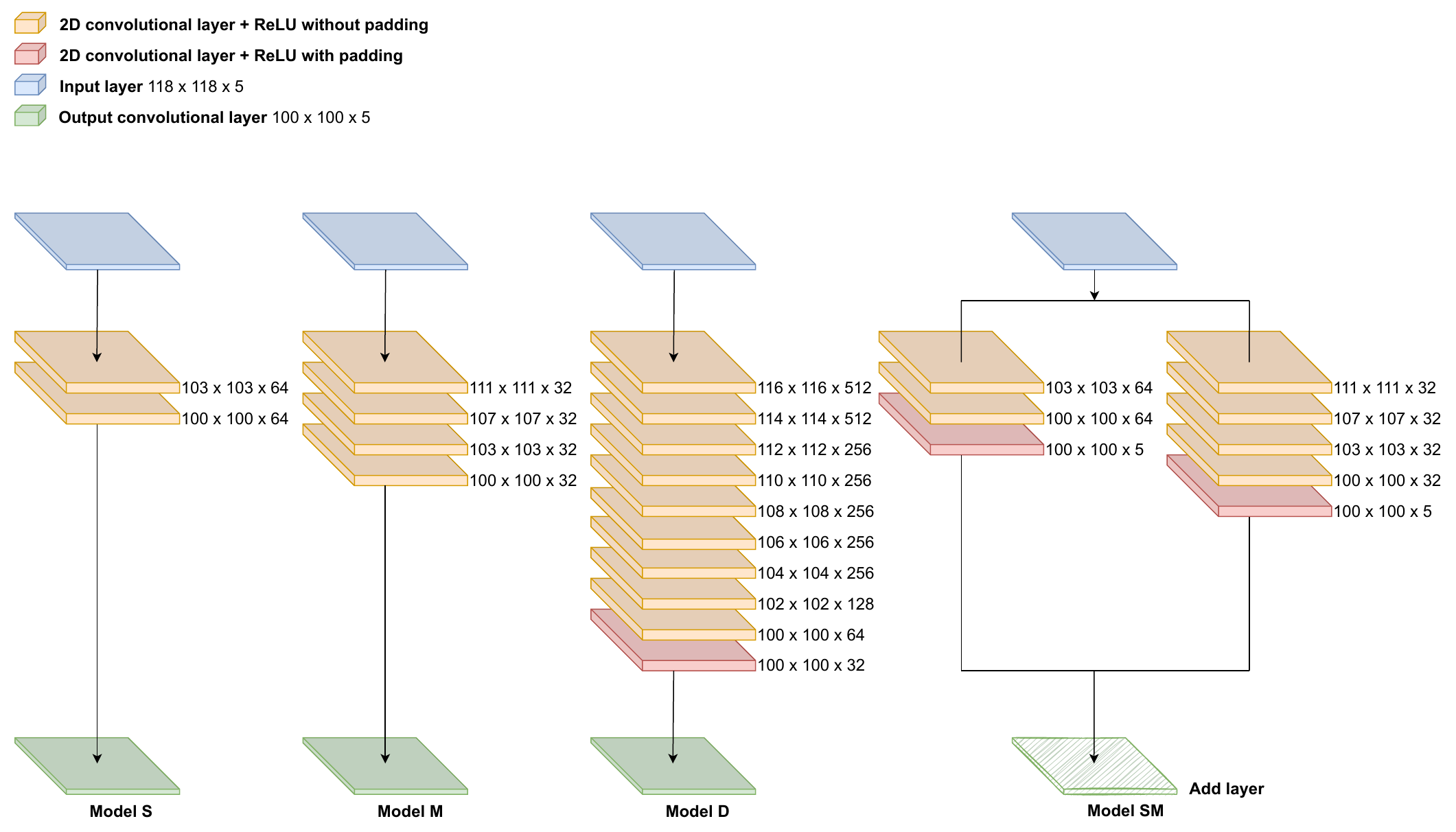}
\caption{Networks with different depths under investigation.}
\label{Fig:Model_SMD}
\end{center}
\end{figure}
In this Section, we study how the network performance depends on the network hyperparameters -- here we focus on the depth of a network and the size of CNN filters.
We take three individual branches from our network of 5 output channels displayed in Fig.~\ref{Fig:CNN_5} as well as combining (some of) them.
 Model S~(S: shallow) shown in Fig.~\ref{Fig:Model_SMD} comes from the shortest branch of the network and  has only two deep layers
while Model D~(D: deep) corresponds to the longest branches with 10 deep layers, and Model M has 4 deep layers.
In addition, Model SM combines Model S and M with the help of layer \texttt{Add}, which simply adds up the contributions from the two models,
as shown in Fig.~\ref{Fig:Model_SMD}. Similarly, Model SMD merges Model S, M and D.

We train all the models with 40896 samples from {\it full cropping}, explained in Section~\ref{sec:net_a}, with extra 4448 samples as the validation set.
The training results are summarized in Table~\ref{tab:net_comp}.
\begin{table}[htp]
 	\centering
	\begin{tabular}{P{5cm}<{\raggedright}  P{5cm} P{7cm} }
		\toprule
		& MSE~(PSNR) in validation& Number of parameters~(million) \\
		\midrule
		Model S & 13.31~(36.89) &  0.15  \\
		Model M  & 11.83~(37.40)  &  0.079 \\ 	
		Model D  & 1.5843~(46.13)  &  6.31 \\
		Model SM  & 7.372~(39.45)  &  0.23 \\
		Model SMD  & 0.8414~(48.88)  &  6.54 \\
		Network in Fig.~\ref{Fig:CNN_5}  & 0.7510~(49.37)  &  6.79 \\
		\bottomrule
	\end{tabular}
	\caption{The comparison of MSE and PSNR in the validation set and the number of trainable parameters~(million) among the different networks.}
	\label{tab:net_comp}
\end{table}
For the models of a single branch, Model D has the best performance due to the largest number of the parameters, followed by Model M and S.
Although Model M is deeper than Model S, it actually has fewer parameters than S but performs better.
That indicates smaller filter sizes are more suitable for the CTIS reconstruction than larger ones.

\begin{figure}[htp]
\begin{center}
\subfloat[]{\raisebox{4.435ex}{\includegraphics[width=0.25\textwidth ]{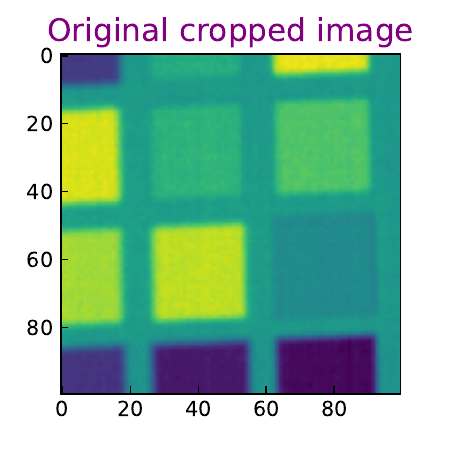}\label{Fig:mod_com_ori}}}
\subfloat[]{\includegraphics[width=0.9\textwidth]{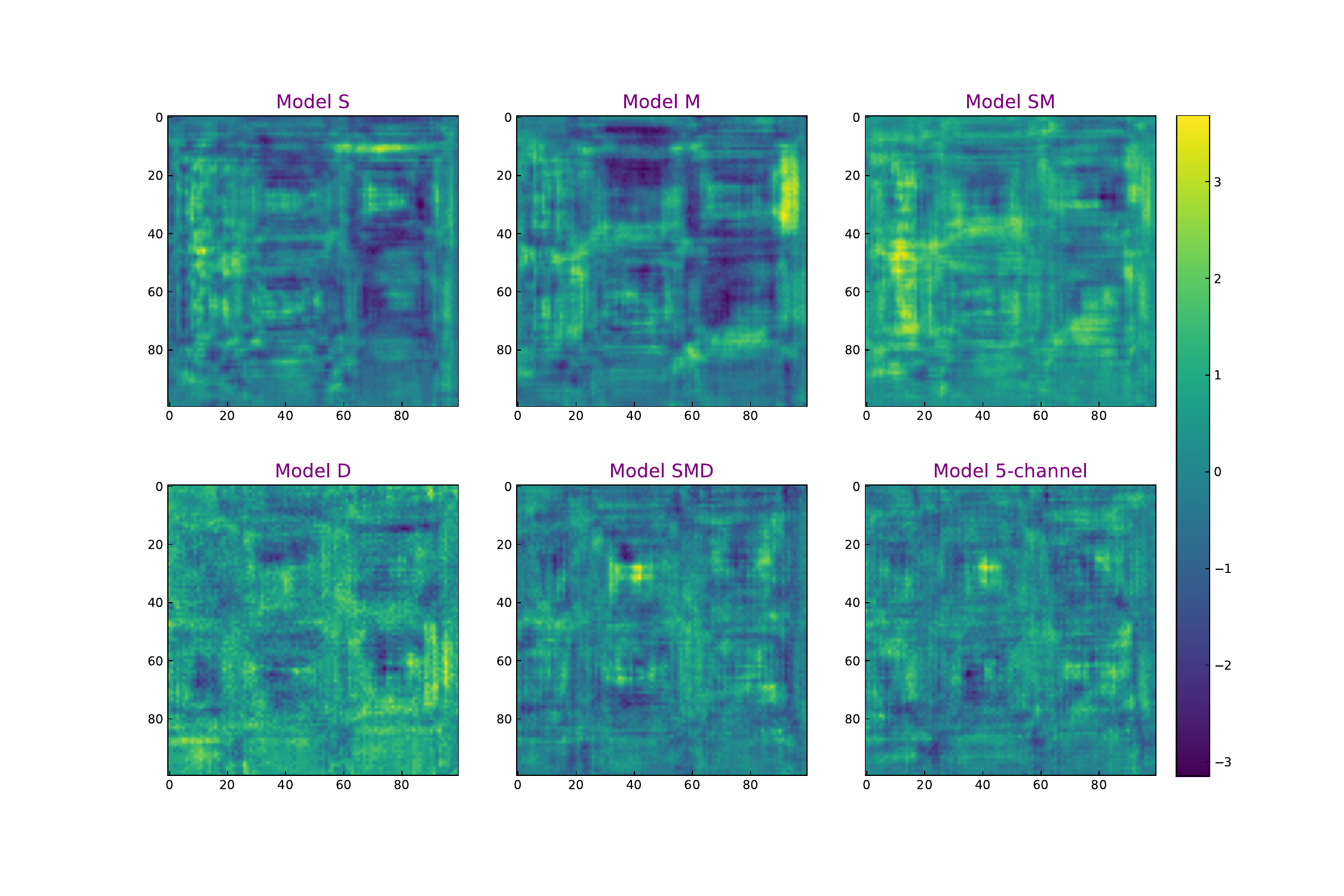}\label{Fig:mod_com_6}}
\caption{{(a) Channel 5 of a sample cube from the validation set.  (b) The corresponding residual error for the models in question. }}
\label{Fig:mod_com}
\end{center}
\end{figure}

On the other hand, Model SM performs noticeably better than Model S and M alone, corroborating the idea of forming a strong learner from multiple weak learners.
That is further reinforced by the MSE reduction from Model D to SMD by nearly a factor of 2 even if the increase on the number of parameters is insignificant.
The improvement from Model SMD to the network with 5 branches is, however, not so dramatic which implies only few of branches are needed to attain decent predictions.

Finally, we present the residual error of a sample from the validation set for the 5th channel in Fig.~\ref{Fig:mod_com_6} with the original image shown in Fig.~\ref{Fig:mod_com_ori}. It is clear that patterns of the error become less visible and the variation of MSE also becomes smaller for models with better performance.

\section{Spectral distortion of reconstructed hyperspectral images}
\label{sec:appSpectral}
\begin{figure}[htp]
	\centering
	
	\subfloat[]{\includegraphics[width=0.45\textwidth]{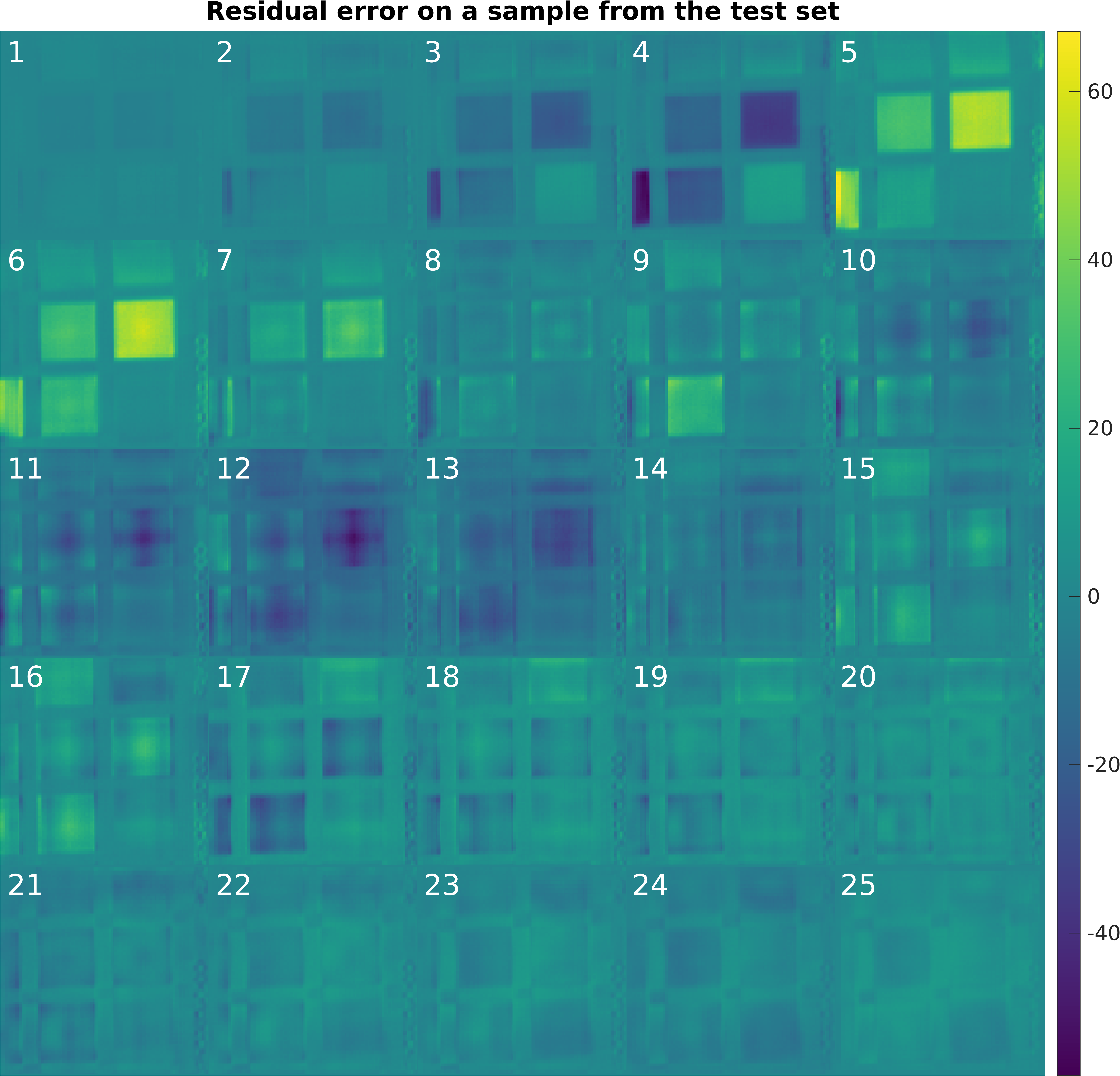}\label{Fig:specDist_mse_a}}
	\hfill
	\subfloat[]{\includegraphics[width=0.45\textwidth]{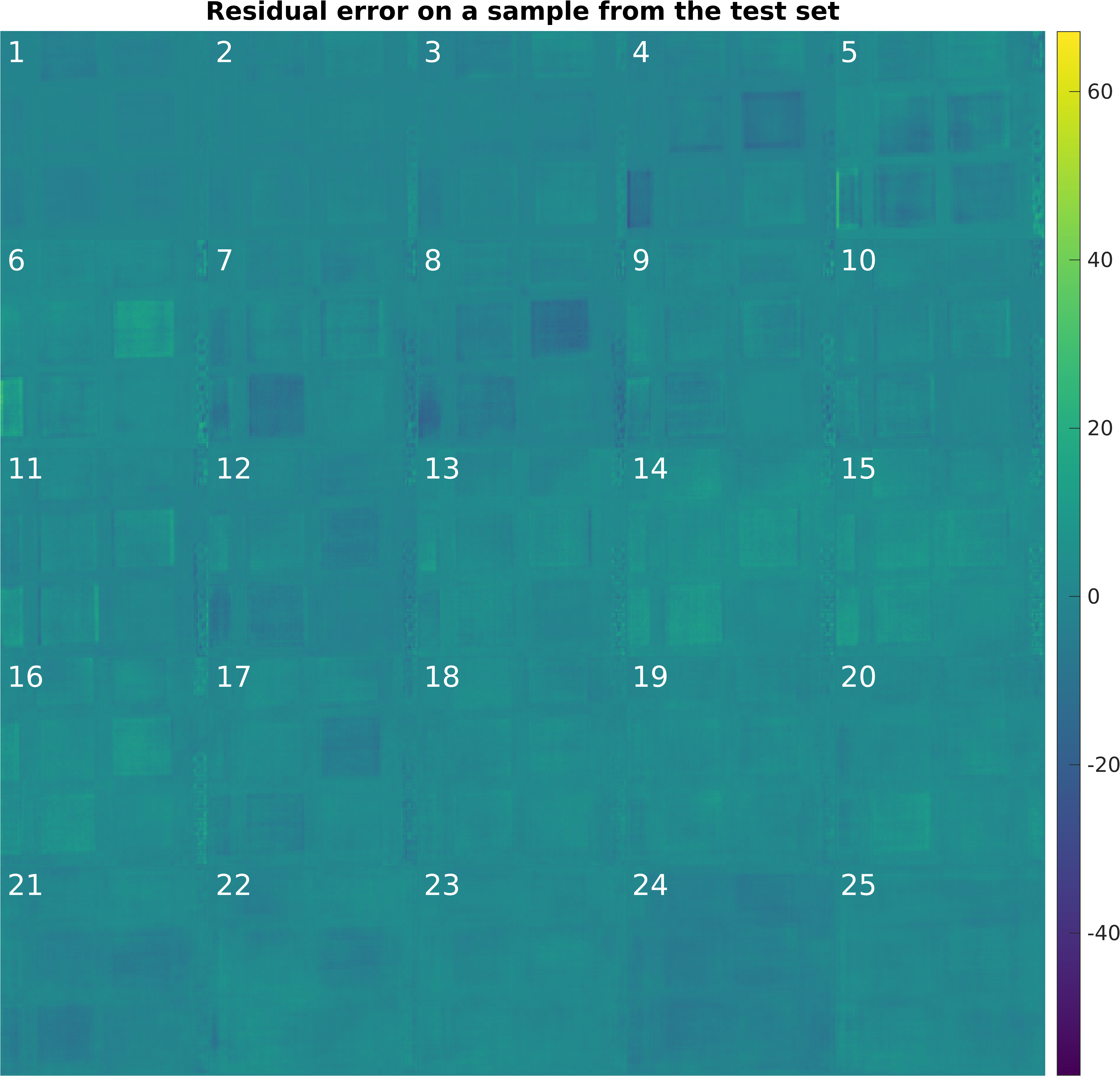}\label{Fig:specDist_mse_b}}
	\hfill
	\caption{The residual error for (a) EM and (b) CNN from a sample in the test set (same sample as in Fig.~\ref{Fig:spec_res_25}) for visualization of the MSE variation among the channels. The same colorbar is used for both (a) and (b) for comparison. The indices for the spectral channels are indicated in white.}
	\label{Fig:specDist_mse}
\end{figure}

Fig.~\ref{Fig:specDist_mse} compares the residual error on the same sample (one used in Fig.~\ref{Fig:spec_res_25}) between (a) EM and (b) CNN, where two colorbars are scaled equally and are identical.
The spectral channel index is specified in the upper-left corner.
Significantly lower residual errors are evident for the CNN reconstruction for all spectral channels. Reconstruction artifacts in the squares of the ColorChecker are clearly present in the EM case.

\end{document}